\documentclass{aa501}
\usepackage{psfig}
\begin{document}
\def\point{$\cdot$ \hspace{0.1cm} $\cdot$ \hspace{0.1cm} $\cdot$}
   \title{The ROSAT Deep Survey}

   \subtitle{VI. X-ray sources and Optical identifications of the Ultra Deep Survey}

   \author{I. Lehmann \inst{1,2}
             \and
           G. Hasinger\inst{1}
             \and
           M. Schmidt \inst{3}
             \and
           R. Giacconi  \inst{4}
             \and
           J. Tr\"umper \inst{5}
             \and
           G. Zamorani \inst{6,7}
             \and
           J.E. Gunn  \inst{8}
             \and
           L. Pozzetti \inst{7}
             \and
           D.P. Schneider  \inst{2}
             \and
           T. Stanke \inst{9}
             \and
           G. Szokoly \inst{1}
             \and
           D. Thompson \inst{3}
             \and 
           G. Wilson \inst{10}
          }

   \offprints{I. Lehmann}

   \institute{Astrophysikalisches Institut Potsdam,
              An der Sternwarte 16, D-14482 Potsdam, Germany\\
              email: ilehmann@aip.de, ghasinger@aip.de, gpszokoly@aip.de
          \and
              Department of Astronomy and Astrophysics, 
              Pennsylvania State University, 
              University Park, PA 16802, USA\\
              email: ingo@astro.psu.edu, dps@astro.psu.edu
          \and
              California Institute of Technology, Pasadena, CA 91125, USA\\
              email: mxs@deimos.caltech.edu,djt@mop.caltech.edu
          \and
              Associated Universities, Inc. 1400 16th Street, NW,
              Suite 730, Washington, DC 20036, USA\\
              email: giacconi@aui.edu
          \and
              Max-Planck-Institute f\"ur extraterrestrische Physik,
              Karl-Schwarzschild-Str. 1, 85748 Garching bei M\"unchen, Germany\\
              email: jtrumper@mpe-garching.mpg.de 
          \and
              Istituto di Radioastronomia del CNR, via Gobetti 101, I-40129,
              Bologna, Italy\\
              email: zamorani@astbo3.bo.astro.it
          \and 
              Osservatorio Astronomico, Via Ranzani 1, I-40127, Bologna, Italy\\
              email: lucia@bo.astro.it
          \and 
              Princeton University Observatory, Peyton Hall, Princeton,
              NJ 08540, USA\\
              email: jeg@astro.princeton.edu
          \and 
              Max-Planck-Institute f\"ur Radioastronomie,
              Auf dem H\"ugel 69, D-53010 Bonn, Germany\\
              email: tstanke@mpifr-bonn.mpg.de
          \and 
              Brown University,
              Physics Department, Providence, RI 02912, USA\\
              email: gillian@het.brown.edu
          }

   \date{02.01.2001 ; 21.03.2001}

   \abstract{
The ROSAT Deep Surveys in the direction of the Lockman Hole are the most sensitive X-ray surveys performed with the ROSAT satellite. 
About 70-80\% of the X-ray background has been resolved into discrete sources at a flux limit of $\sim$10$^{-15}$ erg cm$^{-2}$ s$^{-1}$ in the 0.5-2.0 keV energy band. 
A nearly complete optical identification of 
the ROSAT Deep Survey (RDS) has shown that the great majority of sources are AGNs. 
We describe in this paper the ROSAT Ultra Deep Survey (UDS), an extension of the RDS in the Lockman Hole. The Ultra Deep Survey reaches a flux level of 1.2~$\cdot$~10$^{-15}$ erg cm$^{-2}$ s$^{-1}$ in 0.5-2.0 keV energy band, a level $\sim$4.6 times fainter than the RDS. 
We present nearly complete spectroscopic identifications (90\%) of the sample of 94 X-ray sources based on low-resolution Keck spectra.\\ 
The majority of the sources (57) are broad emission line AGNs (type~I), whereas a further 13 AGNs show only narrow emission lines or broad Balmer emission lines with a large Balmer decrement (type II AGNs) indicating significant optical absorption.
The second most abundant class of objects (10) are groups and clusters of galaxies ($\sim$11\%). Further we found five galactic stars and one ''normal'' emission line galaxy. Eight X-ray sources remain spectroscopically unidentified.
We see no evidence for any change in population from the RDS survey to the UDS survey.\\
The photometric redshift determination indicates in three out of the eight sources the presence of an obscured AGN. Their photometric redshifts, assuming that the spectral energy distribution (SED) in the optical/near-infrared is due to stellar processes, are in the range of $1.2 \le z \le 2.7$.  These objects could belong to the long-sought population of type 2 QSOs, which are predicted by the AGN synthesis models of the X-ray background. 
Finally, we discuss the optical and soft X-ray properties of the type I AGN, type II AGN, and groups and clusters of galaxies, and the implications to the X-ray backround.
\thanks{Tables 3 and 4 are also available in electronic form at the CDS via anonymous ftp to cdsarc.u-strasbg.fr (130.79.128.5)
or via http://cdsweb.u-strasbg.fr/cgi-bin/qcat$?$J/A+A/}
      \keywords{surveys -- galaxies: active -- galaxies: clusters: general
                 -- quasars: emission lines -- galaxies: Seyferts
                 -- X-rays: galaxies
               }
               }

\maketitle
%

\section{Introduction}

The X-ray backgound has been a matter of intense study since its discovery
about 40 years ago by Giacconi et al. (1962). Several deep X-ray surveys using
ROSAT, BeppoSAX, ASCA and recently Chandra and XMM-Newton have resolved a large fraction (60-80 \%) of the soft and hard X-ray background into discrete sources (Hasinger et al. \cite{Has98}, Mushotzky et al. \cite{Mu00}, Giacconi et al. \cite{Gia00}, Hasinger et al. \cite{Has01}.) 

The optical/infrared identification of faint X-ray sources from deep surveys is the key to understanding the nature of the X-ray background. A significant advance was the nearly complete optical identification of the ROSAT Deep Survey (RDS), which contains a sample of 50 PSPC X-ray sources with fluxes above 5.5~$\cdot$~10$^{-15}$ erg cm$^{-2}$ s$^{-1}$ in the 0.5-2.0 keV energy band (Hasinger et al. \cite{Has98}, hereafter Paper I). The spectroscopy has revealed that $\sim$80\% of the optical counterparts are AGNs (Schmidt et al. \cite{Schm98}, hereafter Paper II). 

The very red colour ($R-K^{\prime}>5.0$) of the only two optically unidentified RDS sources indicates either high redshift clusters of galaxies ($z > 1.0$) or obscured AGNs (Lehmann et al. \cite{Leh00a}, hereafter Paper III). This study found a much larger fraction of AGNs than did any previous X--ray survey (Boyle et al. \cite{Boy95}, Georgantopoulos et al. \cite{Geor96}, Bower et al. \cite{Bow96} and McHardy et al. \cite{Mc98}).

Several deep hard X-ray surveys ($>$2 keV) have been started with Chandra and XMM-Newton to date (e.g., Brandt et al. \cite{Bra00}, Mushotzky et al. \cite{Mu00}, Hornschemeier et al., \cite{Horn00}, Giacconi et al. \cite{Gia00} and Hasinger et al. \cite{Has01}), but most of their optical identification is at an early stage or their total number of sources is still relatively small (Barger et al. \cite{Ba01}).

In this paper we present the X-ray sources and their spectroscopic identification of the Ultra Deep Survey (UDS) in the region of the Lockman Hole. The UDS contains 94 X-ray sources with 0.5-2.0 keV fluxes larger than $1.2 \cdot 10^{-15}$~erg~cm$^{-2}$~s$^{-1}$, which is about 4.6 times fainter than our previously optically identified RDS survey. 

The scope of the paper is as follows.
In Sect. 2 we define the X-ray sample and present its X-ray properties. The optical imaging, photometry and spectroscopy of their optical counterparts are described in Sect. 3. 
The optical identification of the new X-ray sources and the catalogue of optical counterparts of the 94 X-ray sources are presented in Sect. 4. 
A $K^{\prime}$ survey covering half of the X-ray survey area is presented in Sect. 5. The photometric redshift determination of three very red sources ($R-K^{\prime} > 5.0$) is shown in Sect. 6.
The implications of our results with respect to the X-ray background are discussed in Sect. 7.

Throughout the paper, we use $H_{\rm 0} = 50$ km s$^{-1}$ Mpc$^{-1}$, $q_{\rm 0} = 0.5$, and $\Lambda = 0$.

\section{The X-ray observations}

The ROSAT Deep Surveys consist of ROSAT PSPC and HRI observations in the direction of the Lockman Hole, a region of an extremely low galactic Hydrogen 
column density  $N_{\rm H}=5.7 \times 10^{19}$ cm$^{-2}$ (Lockman et al. \cite{Loc86}). 
About 207 ksec were accumulated with the PSPC detector (Pfef\-fermann \& Briel \cite{Pfef82}) centered at the direction $\alpha_{2000}=10^{h}52^{m}00^{s},
\delta_{2000}=57^{o}21^{\prime}36''$. The most sensitive area of the PSPC field of view is within a radius of $\sim$20 arcmin from its center. 

The ROSAT PSPC image is the basis for the ROSAT Deep Survey (RDS), which includes a statistically complete sample of 50 X--ray sources with 0.5--2.0 keV fluxes larger than 1.1~$\cdot$~10$^{-14}$ erg cm$^{-2}$ s$^{-1}$ at off-axis angles smaller than 18.5 arcmin and fluxes brighter than 5.5~$\cdot$~10$^{-15}$ erg cm$^{-2}$ s$^{-1}$ at off-axis angles smaller than 12.5 arcmin (see Paper I).

\begin{figure}[h]
\begin{center}
\begin{minipage}{87mm}
\psfig{figure=rosat_deep.ps,bbllx=126pt,bblly=301pt,bburx=520pt,bbury=714pt,width=86mm,clip=}
\end{minipage}
\end{center}
\caption[]{ROSAT PSPC/HRI false colour image of the Lockman Hole region. The HRI sources are shown in green. Red and blue colours indicate PSPC sources in the 0.1--0.5 keV and 0.5--2.0 keV energy bands, respectively. The field size is $\sim$30 arcmin. North is up, east to the left.}
 \label{UDS}
\end{figure}

A pointing of 1112 ksec (net exposure time) with the ROSAT HRI (David et al. \cite{Dav96}) has been obtained centered at the direction $\alpha_{2000}=10^{h}52^{m}43^{s}, \delta_{2000}=57^{o}28^{\prime}48''$. The HRI pointing is shifted about 10 arcmin to the North-East of the PSPC center (see Fig 1. in Paper I). This shift was chosen to allow a more accurate
wobble correction to increase the sensitivity and spatial resolution using some bright optically identified X-ray sources. 

The HRI field of view is about $36 \times 36$ arcmin. The main advantage of the HRI compared to the PSPC is the higher angular resolution of the HRI ($\sim$5 arcsec) compared to that of the PSPC ($\sim$25 arcsec), but the HRI has practically no energy resolution. Fig. \ref{UDS} presents the combined PSPC/HRI image of the overlap region
covered by both pointings. 

In addition to the HRI and PSPC pointings we have obtained a raster scan with the HRI for a total 205 ksec exposure time, which covers nearly the entire field of the PSPC pointing.

   \begin{table}[t]
      \caption{UDS sample properties.}
      \label{UDS_sample}
      \begin{flushleft}
    \begin{tabular}{lcccc}
      \hline\noalign{\smallskip}
Sample & Off-axis angle$^{a}$ & Area & S$_{\rm lim}^{b}$ & N  \\
         & [arcmin]   & [deg$^{2}$]  & [10$^{-15}$ cgs]  &         \\
      \noalign{\smallskip}
      \hspace{0.4cm}(1) & \hspace{0.1cm}(2) & (3) & (4) & (5)   \\\noalign{\smallskip}
      \hline \noalign{\smallskip}
HRI  (H)    & $\le$ 12.0           & 0.126 & 1.2 &  68  \\
PSPC (P1)    & $<$ 12.5$^{c}$     & 0.067 & 5.5 & 14  \\
PSPC (P2)    & 12.5 - 18.5$^{c}$ & 0.118 & 9.6 & 12  \\
       \noalign{\smallskip} \hline \noalign{\smallskip}
      \end{tabular}
      \end{flushleft}
$^{\rm a)}$ The off-axis angle is given with respect to the center of the
pointing of the instrument defining each sample.\\
$^{\rm b)}$ Limiting flux in the 0.5-2.0 keV energy band.\\
$^{\rm c)}$ Excluding the area already included in the HRI sample.
      \end{table}

 The 1112 ksec HRI exposure is the main basis for the Ultra Deep ROSAT Survey (UDS) in the Lockman Hole region. The sample selection and the complete X-ray catalogue are presented in the following sections. For a detailed description of the X-ray observations, the detection algorithm, the astrometric correction and the verification of the analysis procedure by simulation, see Paper I.

\subsection{Sample definition}

The Ultra Deep Survey sample combines three statistically independent samples of HRI and of PSPC sources, whose defining characteristics, both in terms of area and flux limits, are given in Table \ref{UDS_sample}. The total number of sources is 94. Forty-seven of the 68 sources in the HRI sample are also detected with the PSPC, while seven of the 26 sources in the PSPC samples are also detected with the HRI (but outside of the central HRI survey region). For the 54 sources both detected with the HRI and the PSPC we adopt the flux (see Table \ref{xray_tab}) measured with the instrument which defines the sample to which each source belongs. This corresponds to assuming that each sample is defined at the epoch at which its observation has been obtained.

The HRI detector has practically no energy resolution in the 0.1-2.4 keV band, which leads to a significant model-dependence in the count rate to flux conversion.
The conversion of the PSPC count rates in the PSPC hard band (PI channels 52-201) to fluxes in the 0.5-2.0 keV band is straightforward in comparison to the HRI because of the similar band passes (see Paper I). 

The HRI fluxes in the 0.5-2.0 keV band were determined from the HRI count rates using the corresponding exposure time, a vignetting correction, an energy-to-flux conversion factor (ECF) and a point spread function-loss factor determined from simulations (see Table \ref{ECF}). To derive the ECF factor we have assumed a power law spectrum with photon index 2 and galactic absorption (using N$_{\rm H}= 5.7 \times 10^{19}$ cm$^{-2}$), folded through the instrument response. 

   \begin{table}[t]
      \caption{Energy bands and correction factors.}
      \label{ECF}
      \begin{flushleft}
    \begin{tabular}{lccccc}
      \hline\noalign{\smallskip}
Detector & Band & Energy & Pulseheight & ECF$^{a}$ & PCF$^{b}$ \\
         &      & [kev]  & Channels    &     &     \\
      \noalign{\smallskip}
      \hspace{0.4cm}(1) & \hspace{0.1cm}(2) & (3) & (4) & (5)  & (6) \\\noalign{\smallskip}
      \hline \noalign{\smallskip}
PSPC     &  H   & 0.5-2.0 & 52-201 & 0.836  & 0.90  \\
PSPC     &  S   & 0.1-0.4 & 11-41  & 1.519  & 0.90  \\
HRI      &      & 0.1-2.4 & 1-9    & 0.586  & 0.92 \\
       \noalign{\smallskip} \hline \noalign{\smallskip}
      \end{tabular}
      \end{flushleft}
$^{\rm a)}$ Energy to counts conversion factor in cts s$^{-1}$ for a source with 0.5-2.0 keV flux of 10$^{-11}$ erg cm$^{-2}$ s$^{-1}$\\
$^{\rm b)}$ PSF loss correction factor from simulations (see Paper I).\\
      \end{table}

\subsection{The X-ray source catalogue}

Table \ref{xray_tab} provides the complete catalogue of the 94 X-ray sources from the Ultra Deep ROSAT Survey in the Lockman Hole region. The first two columns give the source name and an internal source number. The capital letters in parenthesis mark whether the source belongs to the HRI sample (H) or to the PSPC samples (P1 or P2). The weighted coordinates of the X-ray sources for an
equinox of J2000.0 are shown in columns 3 and 4. The 1$\sigma$ error of their position (in arcsec), including statistical and systematic errors, is given in column 5. The capital P indicates that the X-ray position is based on the 207 ksec PSPC exposure. The positions based on the 1112 ksec HRI pointing and the 205 ksec HRI raster scan are marked with capital letters H and R, respectively.
Columns 6 and 7 show the distance of the sources from the center of the
HRI and PSPC field (in arcmin). Columns 8 and 9 contain the 0.5-2.0 keV HRI and PSPC fluxes of the sources in units of 10$^{-14}$ erg~cm$^{-2}$~s$^{-1}$. The PSPC flux marked with parenthesis is lower than the flux limits of the PSPC samples (see Table \ref{UDS_sample}). These sources would not belong to the PSPC samples without HRI detection. The hardness ratio HR1 $=$ (H-S)/(H$+$S) derived from the PSPC hard and soft bands (see Table \ref{ECF}) is given in column 10.

\subsection{The differences between the UDS and the RDS samples}

The UDS sample contains all X-ray sources from the RDS sample (see Paper I and II),  except the sources 36 and 116,
 which are in the HRI area (HRI off-axis angle smaller than 12 arcmin) and have not been detected by the HRI.
The re-analysis of the PSPC data has led to slightly different
%
%
   \begin{table*}[h]
     \caption{X-ray source catalogue.}
      \label{xray_tab}
      \begin{flushleft}
      \begin{tabular}{lrcccccrrr}
      \hline\noalign{\smallskip}
Source name & $Nr.$\hspace{0.3cm} & $\alpha_{2000}$ & $\delta_{2000}$ & Err. & HRI & PSPC  & $f_{\rm HRI}$\hspace{0.5cm} & $f_{\rm PSPC}$\hspace{0.5cm} & HR1\hspace*{0.5cm}\\ 
 &  &  &  &  &  \multicolumn{2}{c}{Off-axis angle} & \multicolumn{2}{c}{[0.5-2.0 keV]} & \\ \noalign{\smallskip}
\hspace*{0.7cm}(1) & (2)\hspace{0.3cm} & (3) & (4) & (5) &  (6) & (7) & (8)\hspace{0.5cm} & (9)\hspace{0.5cm} & (10)\hspace*{0.5cm}\\ \noalign{\smallskip}
      \hline\noalign{\smallskip}
      \noalign{\smallskip}
RX J105230.3$+$573914&  2 (H)   & 10 52 30.3 & 57 39 13.8 & 1.3H & 10.6 & 16.9 & 1.05$\pm$0.08 & 1.16$\pm$0.15  &-0.03$\pm$0.17\\
RX J105302.6$+$573759&  5 (H)   & 10 53 02.6 & 57 37 58.8 & 1.2H &  9.5 & 17.1 & 1.00$\pm$0.06 & (0.70$\pm$0.11)&-0.30$\pm$0.10\\
RX J105316.8$+$573552&  6 (H)   & 10 53 16.8 & 57 35 52.4 & 0.9H &  8.4 & 16.3 & 9.32$\pm$0.15 &13.23$\pm$0.55  &-0.29$\pm$0.02\\
RX J105131.1$+$573440&  8 (H)   & 10 51 31.1 & 57 34 40.4 & 0.8H & 11.3 & 13.0 &12.14$\pm$0.16 &11.78$\pm$0.27  &-0.24$\pm$0.02\\ 
RX J105154.4$+$573438&  9 (H)   & 10 51 54.4 & 57 34 38.0 & 1.0H &  8.8 & 12.1 & 1.09$\pm$0.07 & 1.93$\pm$0.12  &-0.23$\pm$0.06\\
\noalign{\smallskip}
RX J105108.4$+$573345&  11 (P2) & 10 51 08.4 & 57 33 45.4 & 1.6H & 13.7 & 13.6 & 1.04$\pm$0.09 & 1.29$\pm$0.12  &0.18$\pm$0.12\\
RX J105149.0$+$573249&  12 (H)  & 10 51 49.0 & 57 32 48.6 & 1.0H &  8.3 & 10.5 & 0.72$\pm$0.07 & 1.65$\pm$0.13  &1.00$\pm$0.78\\
RX J105213.3$+$573222&  13 (H)  & 10 52 13.3 & 57 32 21.6 & 1.1H &  5.4 &  9.8 & 0.60$\pm$0.05 & 0.61$\pm$0.07  &-0.02$\pm$0.16\\
RX J105242.5$+$573159&  14 (H)  & 10 52 42.5 & 57 31 59.2 & 1.3H &  3.2 & 10.5 & 0.30$\pm$0.04 & 0.76$\pm$0.08  &0.72$\pm$0.52\\
RX J105300.0$+$573155&  15 (H)  & 10 53 00.0 & 57 31 55.1 & 1.3H &  3.8 & 11.7 & 0.34$\pm$0.04 & (0.35$\pm$0.07)&-0.33$\pm$0.22\\
\noalign{\smallskip}
RX J105339.7$+$573105&  16 (H)  & 10 53 39.7 & 57 31 05.3 & 0.9H &  7.9 & 15.1 & 4.07$\pm$0.09 & 3.72$\pm$0.20  &-0.59$\pm$0.02\\
RX J105104.2$+$573054&  17 (P1) & 10 51 04.2 & 57 30 53.7 & 1.6H & 13.6 & 11.8 & 0.62$\pm$0.07 & 0.69$\pm$0.10  &0.11$\pm$0.19\\
RX J105228.4$+$573104&  18 (H)  & 10 52 28.4 & 57 31 04.4 & 1.2H &  3.0 &  8.9 & 0.26$\pm$0.04 & (0.21$\pm$0.05)&-0.45$\pm$0.19\\
RX J105137.4$+$573044&  19 (H)  & 10 51 37.4 & 57 30 44.4 & 1.0H &  9.1 &  9.0 & 0.88$\pm$0.06 & 1.08$\pm$0.10  &-0.24$\pm$0.07\\
RX J105410.3$+$573039&  20 (H)  & 10 54 10.3 & 57 30 39.3 & 1.4H & 11.9 & 18.5 & 1.10$\pm$0.08 & 2.26$\pm$0.15  &0.47$\pm$0.11\\
\noalign{\smallskip}
RX J105224.7$+$573010&  23 (H)  & 10 52 24.7 & 57 30 10.2 & 1.5H &  2.8 &  7.9 & 0.26$\pm$0.03 & 0.60$\pm$0.09  &0.02$\pm$0.20\\
RX J105044.4$+$572922&  24 (P2) & 10 50 44.4 & 57 29 21.7 & 4.7P & 16.0 & 13.0 & -             & 1.00$\pm$0.09  &-0.31$\pm$0.07\\
RX J105344.9$+$572841&  25 (H)  & 10 53 44.9 & 57 28 40.5 & 1.1H &  8.3 & 14.6 & 1.19$\pm$0.07 & 1.76$\pm$0.14  &-0.09$\pm$0.07\\
RX J105020.3$+$572808&  26 (P2) & 10 50 20.3 & 57 28 07.8 & 6.5P & 19.2 & 15.3 & -             & 0.97$\pm$0.10  &-0.06$\pm$0.14\\
RX J105350.3$+$572710&  27 (H)  & 10 53 50.3 & 57 27 09.6 & 2.0H &  9.2 & 14.7 & 0.51$\pm$0.05 & 1.34$\pm$0.12  &0.16$\pm$0.17\\
\noalign{\smallskip}
RX J105421.1$+$572545&  28 (P2) & 10 54 21.1 & 57 25 44.5 & 1.0H & 13.5 & 18.4 &17.36$\pm$0.21 &21.54$\pm$0.43  &1.00$\pm$0.06\\
RX J105335.1$+$572542&  29 (H)  & 10 53 35.1 & 57 25 42.4 & 0.8H &  7.6 & 12.3 & 2.17$\pm$0.07 & 5.14$\pm$0.18  &-0.34$\pm$0.03\\
RX J105257.1$+$572507&  30 (H)  & 10 52 57.1 & 57 25 07.2 & 0.9H &  4.1 &  7.2 & 0.78$\pm$0.05 & 0.81$\pm$0.10  &-0.53$\pm$0.05\\
RX J105331.8$+$572454&  31 (H)  & 10 53 31.8 & 57 24 53.9 & 0.9H &  7.6 & 11.7 & 2.30$\pm$0.09 & 3.41$\pm$0.16  &-0.18$\pm$0.04\\
RX J105239.7$+$572432&  32 (H)  & 10 52 39.7 & 57 24 31.7 & 0.8H &  4.3 &  4.8 & 7.02$\pm$0.15 & 6.88$\pm$0.21  &-0.55$\pm$0.02\\
\noalign{\smallskip}
RX J105200.0$+$572424&  33 (H)  & 10 52 00.0 & 57 24 24.5 & 1.9H &  7.3 &  2.0 & 0.22$\pm$0.04 & (0.41$\pm$0.07)&-0.28$\pm$0.15\\
RX J105258.4$+$572356&  34 (H)  & 10 52 58.4 & 57 23 55.8 & 1.6H &  5.3 &  7.1 & 0.24$\pm$0.03 & (0.38$\pm$0.08)&-0.25$\pm$0.26\\
RX J105039.7$+$572335&  35 (P1) & 10 50 39.7 & 57 23 35.1 & 1.8R & 17.4 & 11.8 & -             & 1.59$\pm$0.11  &-0.29$\pm$0.06\\ 
RX J105247.9$+$572116&  37 (H)  & 10 52 47.9 & 57 21 16.3 & 0.8H &  7.6 &  5.7 & 2.77$\pm$0.10 & 2.55$\pm$0.17  &-0.73$\pm$0.02\\
RX J105329.2$+$572104&  38 (H)  & 10 53 29.2 & 57 21 03.7 & 1.3H &  9.9 & 11.3 & 0.86$\pm$0.06 & 0.80$\pm$0.09  &-0.30$\pm$0.08\\
\noalign{\smallskip}
RX J105209.5$+$572104&  39 (H)  & 10 52 09.5 & 57 21 04.4 & 1.8H &  9.0 &  1.6 & 0.26$\pm$0.06 & (0.43$\pm$0.06)&-0.42$\pm$0.10\\
RX J105318.1$+$572042&  41 (H)  & 10 53 18.1 & 57 20 42.0 & 2.4H &  9.4 &  9.9 & 0.27$\pm$0.04 & 2.15$\pm$0.29  &0.03$\pm$0.10\\
RX J105015.6$+$572000&  42 (P2) & 10 50 15.6 & 57 20 00.2 & 4.7P & 21.7 & 15.2 & -             & 1.52$\pm$0.12  &-0.36$\pm$0.06\\
RX J105105.2$+$571924&  43 (P1) & 10 51 05.2 & 57 19 23.9 & 1.9R & 16.8 &  8.9 & -             & 0.99$\pm$0.10  &-0.28$\pm$0.07\\
RX J105319.0$+$571852&  45 (H)  & 10 53 19.0 & 57 18 51.9 & 1.5H & 11.1 & 10.5 & 0.63$\pm$0.06 & 1.46$\pm$0.09  &0.02$\pm$0.12\\
\noalign{\smallskip}
RX J105120.2$+$571849&  46 (P1) & 10 51 20.2 & 57 18 49.2 & 1.9R & 15.1 &  7.3 & 0.58$\pm$0.09 & 1.03$\pm$0.11  &-0.18$\pm$0.09\\
RX J105244.4$+$571732&  47 (H)  & 10 52 44.4 & 57 17 31.9 & 1.6H & 11.3 &  7.2 & 0.51$\pm$0.05 & 0.71$\pm$0.07  &-0.35$\pm$0.08\\
RX J105046.1$+$571733&  48 (P1) & 10 50 46.1 & 57 17 32.8 & 1.8R & 19.4 & 12.0 & -             & 1.73$\pm$0.12  &-0.08$\pm$0.08\\
RX J105117.0$+$571554&  51 (P1) & 10 51 17.0 & 57 15 54.1 & 5.1P & 17.4 &  9.5 & -             & 0.85$\pm$0.12  &0.04$\pm$0.22\\
RX J105243.1$+$571544&  52 (P1) & 10 52 43.1 & 57 15 44.4 & 1.2H & 13.0 &  8.5 & 1.11$\pm$0.09 & 1.43$\pm$0.10  &-0.36$\pm$0.06\\
\noalign{\smallskip}
RX J105206.0$+$571529&  53 (P1) & 10 52 06.0 & 57 15 28.7 & 4.6P & 14.3 &  7.2 & 0.43$\pm$0.08 & 0.55$\pm$0.08  &-0.15$\pm$0.12\\
RX J105307.2$+$571506&  54 (P1) & 10 53 07.2 & 57 15 05.6 & 1.9H & 14.1 & 11.1 & 0.79$\pm$0.08 & 1.20$\pm$0.11  &0.05$\pm$0.10\\
RX J105009.3$+$571443&  55 (P2) & 10 50 09.3 & 57 14 42.8 & 6.6P & 25.1 & 17.7 & -             & 1.03$\pm$0.12  &0.02$\pm$0.06\\
RX J105020.2$+$571423&  56 (P2) & 10 50 20.2 & 57 14 22.8 & 1.8R & 24.1 & 16.5 & -             & 3.41$\pm$0.17  &-0.44$\pm$0.02\\
RX J105237.9$+$571254&  58 (P1) & 10 52 37.9 & 57 12 53.5 & 6.2P & 15.9 & 10.6 & -             & 0.58$\pm$0.08  &0.07$\pm$0.21\\
\noalign{\smallskip}
RX J105324.6$+$571236&  59 (P2) & 10 53 24.6 & 57 12 35.7 & 2.8P & 17.2 & 14.6 & -             & 1.41$\pm$0.11  &0.27$\pm$0.14\\ 
RX J105248.4$+$571203&  60 (P1) & 10 52 48.4 & 57 12 02.7 & 5.0P & 16.8 & 12.0 & -             & 0.67$\pm$0.08  &-0.00$\pm$0.19\\
RX J105127.0$+$571129&  61 (P1) & 10 51 27.0 & 57 11 29.0 & 6.0P & 20.1 & 12.4 & -             & 0.64$\pm$0.09  &-0.07$\pm$0.14\\
RX J105201.5$+$571044&  62 (P1) & 10 52 01.5 & 57 10 44.2 & 1.6R & 18.9 & 11.9 & -             & 3.49$\pm$0.19  &-0.31$\pm$0.03\\
RX J105055.3$+$570652&  67 (P2) & 10 50 55.3 & 57 06 51.9 & 8.1P & 26.3 & 18.5 & -             & 1.21$\pm$0.14  &0.23$\pm$0.16\\
      \noalign{\smallskip}
      \hline
      \end{tabular}
      \end{flushleft}
   \end{table*}
\addtocounter{table}{-1}

%
%
   \begin{table*}[t]
     \caption{(continued).}
      \begin{flushleft}
      \begin{tabular}{lrcccccrrr}
      \hline\noalign{\smallskip}
Source name & $Nr.$\hspace{0.3cm} & $\alpha_{2000}$ & $\delta_{2000}$ & Err. & HRI & PSPC  & $f_{\rm HRI}$\hspace{0.5cm} & $f_{\rm PSPC}$\hspace{0.5cm} & HR1\hspace*{0.5cm}\\ 
 &  &  &  &  &  \multicolumn{2}{c}{Off-axis angle} & \multicolumn{2}{c}{[0.5-2.0 keV]} & \\ \noalign{\smallskip}
\hspace*{0.7cm}(1) & (2)\hspace{0.3cm} & (3) & (4) & (5) &  (6) & (7) & (8)\hspace{0.5cm} & (9)\hspace{0.5cm} & (10)\hspace*{0.5cm}\\ \noalign{\smallskip}
      \hline\noalign{\smallskip}
      \noalign{\smallskip}
RX J105215.7$+$570411&  70 (P2)   & 10 52 15.7 & 57 04 11.0 & 5.0P & 24.9 & 18.5 & -             & 1.82$\pm$0.17  &-0.12$\pm$0.09\\
RX J105008.2$+$573135&  73 (P2)   & 10 50 08.2 & 57 31 34.7 & 8.2P & 21.0 & 18.3 & -             & 1.31$\pm$0.14  &0.78$\pm$0.51\\
RX J105125.4$+$573050&  75 (H)    & 10 51 25.4 & 57 30 49.8 & 1.8H & 10.6 &  9.9 & 0.43$\pm$0.06 & 0.59$\pm$0.10  &1.00$\pm$1.45\\
RX J105259.2$+$573031&  77 (H)    & 10 52 59.2 & 57 30 30.8 & 1.1H &  2.8 & 10.6 & 0.83$\pm$0.05 & 0.65$\pm$0.09  &-0.14$\pm$0.13\\
RX J105144.8$+$572808&  80 (H)    & 10 51 44.8 & 57 28 07.7 & 1.6H &  7.9 &  6.2 & 0.21$\pm$0.04 & 0.55$\pm$0.07  &0.24$\pm$0.26\\
\noalign{\smallskip}
RX J105312.4$+$572507&  82 (H)    & 10 53 12.4 & 57 25 06.8 & 1.2H &  5.4 &  9.2 & 0.29$\pm$0.04 & (0.44$\pm$0.08)&0.10$\pm$0.26\\
RX J105217.0$+$572017&  84 (H)    & 10 52 17.0 & 57 20 17.1 & 2.0H &  9.2 &  2.8 & 0.27$\pm$0.05 & 0.67$\pm$0.09  &0.68$\pm$0.71\\
RX J105241.8$+$573651& 104 (H)    & 10 52 41.8 & 57 36 50.8 & 1.3H &  8.0 & 15.0 & 0.52$\pm$0.05 & (0.81$\pm$0.10)&0.39$\pm$0.27\\
RX J105348.7$+$573033& 117 (H)    & 10 53 48.7 & 57 30 33.5 & 1.4H &  9.0 & 15.9 & 0.66$\pm$0.06 & 1.64$\pm$0.14  &1.00$\pm$1.51\\
RX J105309.4$+$572822& 120 (H)    & 10 53 09.4 & 57 28 21.9 & 1.1H &  3.5 & 10.2 & 0.78$\pm$0.05 & (0.47$\pm$0.08)&-0.20$\pm$0.15\\
\noalign{\smallskip}
RX J105350.7$+$572515& 128 (H)    & 10 53 50.7 & 57 25 15.4 & 2.0H &  9.7 & 14.3 & 0.26$\pm$0.04 & -              & -       \\ 
RX J105340.4$+$572349& 131 (H)    & 10 53 40.4 & 57 23 48.6 & 2.7H &  9.2 & 12.7 & 0.18$\pm$0.04 & (0.36$\pm$0.07)&0.94$\pm$9.14:\\
RX J105406.7$+$571339& 151 (P2)   & 10 54 06.7 & 57 13 39.3 &10.6P & 18.9 & 18.5 & -             & 1.03$\pm$0.15  &-0.55$\pm$0.06\\
RX J105341.0$+$573522& 228 (H)    & 10 53 41.0 & 57 35 22.3 & 2.6H & 10.2 & 18.0 & 0.25$\pm$0.05 & 1.68$\pm$0.14  & 0.74$\pm$0.61\\
RX J105346.6$+$573517& 229 (H)    & 10 53 46.6 & 57 35 17.1 & 4.6H & 10.7 & 18.5 & 0.20$\pm$0.05 & -              & -       \\
\noalign{\smallskip}
RX J105336.4$+$573802& 232 (H)    & 10 53 36.4 & 57 38 02.2 & 1.1H & 11.7 & 19.6 & 3.25$\pm$0.13 & 3.08$\pm$0.17  &0.32$\pm$0.12\\ 
RX J105303.9$+$572925& 426 (H)    & 10 53 03.9 & 57 29 24.9 & 1.3H &  2.8 & 10.3 & 0.34$\pm$0.04 & (0.23$\pm$0.06)&1.00$\pm$3.02\\
RX J105324.7$+$572819& 428 (H)    & 10 53 24.7 & 57 28 19.3 & 1.4H &  5.6 & 12.0 & 0.31$\pm$0.05 & (0.39$\pm$0.08)&-0.20$\pm$0.37\\ 
RX J105258.2$+$572249& 434 (H)    & 10 52 58.2 & 57 22 49.4 & 1.7H &  6.3 &  7.0 & 0.18$\pm$0.03 & (0.35$\pm$0.07)&0.02$\pm$1.49\\
RX J105306.2$+$573426& 477 (H)    & 10 53 06.2 & 57 34 25.7 & 1.3H &  6.4 & 14.3 & 0.37$\pm$0.04 & (0.40$\pm$0.10)&-0.37$\pm$0.87\\
\noalign{\smallskip}
RX J105243.4$+$572800& 486 (H)    & 10 52 43.4 & 57 28 00.3 & 1.7H &  0.8 &  7.3 & 0.15$\pm$0.03 & (0.27$\pm$0.05)&1.00$\pm$1.98\\ 
RX J105114.5$+$571616& 504 (P1)   & 10 51 14.5 & 57 16 16.0 & 1.8R & 17.3 &  9.5 & -             & 1.01$\pm$0.11  &-0.38$\pm$0.07\\ 
RX J105254.3$+$572343& 513 (H)    & 10 52 54.3 & 57 23 42.9 & 1.3H &  5.3 &  6.5 & 0.43$\pm$0.05 & (0.35$\pm$0.06)&-0.46$\pm$0.10\\
RX J105220.1$+$572307& 607 (H)    & 10 52 20.1 & 57 23 07.3 & 1.4H &  6.5 &  1.9 & 0.27$\pm$0.05 & (0.36$\pm$0.06)&-0.25$\pm$0.17\\ 
RX J105311.7$+$572306& 634 (H)    & 10 53 11.7 & 57 23 06.1 & 1.8H &  6.9 &  8.8 & 0.20$\pm$0.04 & -              & -   \\
\noalign{\smallskip}
RX J105245.7$+$573748& 801 (H)    & 10 52 45.7 & 57 37 47.9 & 1.9H &  9.0 & 16.0 & 0.31$\pm$0.05 & -              & -       \\
RX J105222.4$+$573737& 802 (H)    & 10 52 22.4 & 57 37 36.8 & 1.6H &  9.2 & 15.1 & 0.42$\pm$0.06 & -              & -       \\ 
RX J105312.5$+$573425& 804 (H)    & 10 53 12.5 & 57 34 24.7 & 1.3H &  6.9 & 14.7 & 0.43$\pm$0.04 & -              & -       \\
RX J105348.2$+$573355& 805 (H)    & 10 53 48.2 & 57 33 54.7 &  4.3H & 10.1 & 17.7 & 0.17$\pm$0.05 & -              & -         \\
RX J105244.7$+$572122& 814 (H)    & 10 52 44.7 & 57 21 22.2 & 1.3H &  7.4 &  5.3 & 0.48$\pm$0.04 & 0.61$\pm$0.16  &0.68$\pm$0.68\\
\noalign{\smallskip}
RX J105329.5$+$573538& 815 (H)    & 10 53 29.5 & 57 35 37.9 & 2.2H &  9.2 & 17.1 & 0.18$\pm$0.04 & -              & -         \\ 
RX J105225.9$+$571905& 817 (H)    & 10 52 25.9 & 57 19 04.8 & 2.4H & 10.0 &  4.4 & 0.23$\pm$0.05 & -              & -       \\
RX J105322.2$+$572852& 821 (H)    & 10 53 22.2 & 57 28 51.6 & 1.5H &  5.2 & 11.9 & 0.25$\pm$0.03 & -              & -       \\
RX J105206.3$+$572214& 825 (H)    & 10 52 06.3 & 57 22 14.0 & 1.5H &  8.2 &  0.4 & 0.33$\pm$0.04 & -              & -       \\ 
RX J105303.7$+$573532& 827 (H)    & 10 53 03.7 & 57 35 31.9 & 2.0H &  7.3 & 15.0 & 0.18$\pm$0.04 & -              & -       \\
\noalign{\smallskip}
RX J105357.1$+$573241& 828 (H)    & 10 53 57.1 & 57 32 41.1 & 2.0H & 10.7 & 18.0 & 0.45$\pm$0.07 & -              & -       \\ 
RX J105207.7$+$573842& 832 (H)    & 10 52 07.7 & 57 38 42.0 & 3.5H & 11.0 & 16.1 & 0.24$\pm$0.07 & -              & -       \\
RX J105137.4$+$572857& 837 (H)    & 10 51 37.4 & 57 28 57.2 & 2.6H &  8.8 &  7.4 & 0.19$\pm$0.04 & -              & -         \\
RX J105244.4$+$574045& 840 (H)    & 10 52 44.4 & 57 40 45.0 & 4.1H & 12.0 & 18.8 & 0.22$\pm$0.03 & -              & -         \\ 
RX J105358.5$+$572925& 861 (H)    & 10 53 58.5 & 57 29 25.4 & 3.2H & 10.1 & 16.5 & 0.18$\pm$0.04 & -              & -         \\
\noalign{\smallskip}
RX J105236.4$+$573748& 866 (H)    & 10 52 36.4 & 57 37 48.0 &  2.9H &  9.0 & 15.7 & 0.18$\pm$0.04 & -              & -       \\
RX J105225.3$+$572246& 870 (H)    & 10 52 25.3 & 57 22 45.5 & 1.4H &  6.5 &  2.5 & 0.36$\pm$0.04 & -              & -       \\ 
RX J105252.9$+$572901& 901 (H)    & 10 52 52.9 & 57 29 00.7 & 2.7H &  1.3 &  8.9 & 0.12$\pm$0.04 & -              & -       \\
RX J105251.2$+$572011& 905 (H)    & 10 52 51.2 & 57 20 11.0 & 2.7H &  8.7 &  6.5 & 0.14$\pm$0.04 & -              & -       \\
      \noalign{\smallskip}
      \hline
      \end{tabular}
      \end{flushleft}
   \end{table*}
\clearpage 

\begin{figure}[h]
     \begin{minipage}{87mm}
        \psfig{figure=fratio_HR1_AGN_rest.ps,bbllx=45pt,bblly=47pt,bburx=498pt,bbury=650pt,width=87mm,angle=-90}
     \end{minipage}
     \caption[]{HRI to PSPC flux ratio in the 0.5-2.0 keV energy band versus the PSPC hardness ratio HR1 for those X-ray sources detected both in the HRI and in the PSPC pointings. The symbols mark different classes of counterparts; filled circles - type I AGNs, open circles -- type II AGNs, and triangles - groups/clusters of galaxies, galaxies, and stars (see Table \ref{id_tab}). 
The error bar in the upper right gives the mean error of the data points. 
The lines show the theoretical HRI to PSPC flux ratio in depenence of 
an increasing Hydrogen column densitiy 
from N$_{\rm H}=0 \times$ 10$^{22}$ cm$^{-2}$ to 1.0 $\times$ 10$^{22}$ cm$^{-2}$ (from left to right) in steps of 0.0057, 0.02, 0.03, 0.04, 0.06, 0.08, 0.1, 0.12, 0.16, 0.3, 0.6, and 0.8 $\times$ 10$^{22}$ cm$^{-2}$ for three different photon indices (dotted line -- $\Gamma=1$, solid line -- $\Gamma=2$, and dashed line -- $\Gamma=3$).} 
               \label{flux_ratio}
\end{figure}
%

\noindent PSPC fluxes compared to Paper I. The new PSPC fluxes are statistically consistent with Paper I fluxes. The flux limit of the PSPC subsample P2 (see Table \ref{UDS_sample}) is marginally lower than that limit for the same off-axis angles in the RDS sample (1.1 $\times$ 10$^{-14}$ erg~cm$^{-2}$~s$^{-1}$). 

The PSPC fluxes differ from the HRI fluxes for several sources (see Table \ref{xray_tab}). Fig. \ref{flux_ratio} shows the HRI to PSPC flux ratio plotted versus the PSPC hardness ratio HR1 for all sources detected both with the HRI and the PSPC. The flux ratio of the softer X-ray sources (HR1 $\le -0.5$) scatters around one, whereas the HRI flux of the hard sources (HR1 $\ge$ 0.5)
is clearly lower than the PSPC flux, indicated by an average flux ratio of 0.6.

All spectroscopically identified type II AGNs including the very red sources (see Sect. 4 and 5) are relatively hard sources. 
Their lower HRI to PSPC flux ratio can be due to a harder spectrum than the assumed unabsorbed power-law spectrum with photon index 2, leading to lower HRI fluxes compared to PSPC fluxes for intrinsically absorbed sources. 
We have performed simulations with XSPEC to determine the influence of intrinsic absorption to the HRI to PSPC flux ratio using three different photon indices ($\Gamma = 1, 2, 3$). The models predict a decrease of the flux ratio ( $\le$ 0.8) at apparent column densities (i.e., not corrected for redshift) larger than 1.2 $\times$ 10$^{21}$ cm$^{-2}$ (see Fig. \ref{flux_ratio}).
The lower HRI to PSPC flux ratio of the harder UDS sources (HR1 $\ge$ 0.5) results probably from intrinsic absorption. 

The ASCA Deep Survey in the Lockman Hole by 
Ishisaki et al. (\cite{Ish01}) has confirmed that the X-ray luminous UDS type II AGN, detected with ASCA in the 1-7 keV energy band, show harder spectra than type I AGN, which can be explained by intrinsic absorption with $N_{\rm H} \sim 10^{22-23}$ cm$^{-2}$.

Several groups and clusters of galaxies (e.g., 41C and 228Z) show significant lower HRI fluxes compared to their PSPC fluxes. This is probably due to the lower
sensitivity of the HRI to detect extended X-ray emission. The 0.5-2.0 keV X-ray flux of the UDS groups and clusters of galaxies can be at least a factor of 2 larger than those values, which are based on the HRI, given in Table \ref{xray_tab}.

However, the difference between the PSPC and HRI fluxes for several sources can be also due to X-ray variability, because the fluxes are defined by the epoch of the observation. 

\section{Optical spectroscopy}

Optical images and optical spectra of 50 of the 94 UDS sources are published in Paper III and in Hasinger et al. (\cite{Has99b}, hereafter Paper IV). Finding charts and, when available,
   optical spectra are given in Fig. \ref{spec1} for the remaining 44 objects.

The optical identifications of most of the new X-ray sources is based on their highly accurate HRI positions and Keck $R$ band images with a
limiting magnitude ranging from 23.5 to 26.0. We have found for nearly all sources only one optical counterpart inside the HRI error circle (see Fig. \ref{spec1}). In some cases the HRI detection of PSPC sources (e.g., 15 and 18) is essential to identify the correct optical counterpart. For the new sources, presented here for the first time, three (24, 70, and 151) are only covered by the PSPC exposure. For these sources, due to the lower positional accuracy of the PSPC detector ($25''$), we considered all objects inside the PSPC error circle as possible optical counterparts. For sources 5, 24, 70, 131, 151, 477, 804, 827, 832, and 840 we used Palomar 5-m $V$ band images taken with the 4-Shooter (Gunn et al. \cite{Gun87}) to locate the optical counterparts, as for these cases Keck $R$ images are not available. 

The optical imaging and photometry of the X-ray sources has been previously described in Papers II and III. However, we have to complete the $R$-band coverage (not the entire UDS survey area is covered by Keck LRIS frames).
The optical photometry should still be regarded as somewhat uncertain for those
objects, not covered by the Keck $R$ images.
For instance, the $R$ magnitudes of the objects 9A, 17A, 29A, and 54A are significant lower than those given in Paper II due to the more reliable Keck photometry compared to the $R$ band photometry based on imaging with the University of Hawaii 2.2-m telescope (see Paper II). A more complete and reliable photometry will be published in a future paper.

Optical spectroscopy of nearly all new UDS sources was performed with the Low Resolution Imaging Spectrometer (LRIS; Oke et al. \cite{Oke95}) at the Keck I and II 10-m 
telescopes during observing runs in February and December 1995, April 1996, April 1997, March 1998 and March 1999. The spectra were taken either through multislit masks using 1.4$''$ and 1.5$''$ wide slits or through 0.7$''$ and 1.0$''$ wide long slits. The detector is a back-illuminated 2048 $\times$ 2048 Tektronics CCD. A 300 lines~mm$^{-1}$ grating produces a spectral resolution of $\sim$10 \AA~for the long slit spectra and of $\sim$15~\AA~for the mask spectra. The wavelength coverage of the spectra varies within 3800-8900 \AA. In addition, long slit spectra of some relatively bright counterparts ($R<21$) were obtained with LRIS at the Keck II telescope during service observing in February, March and April 1998. 
The exposure times for long slit spectra range from 120 to 3600 sec, for mask spectra from 1800 to 3600 sec.

The optical spectrum of the bright star 232A ($R=11.7$) was taken in March 1998 at the Calar Alto 3.5-m telescope using the Boller \& Chivens Cass Twin spectrograph. The detector is a 800 $\times$ 2000 SITe CCD. A 600 lines mm$^{-1}$ transmission grating and a slit width of 1.5$''$ produce spectra from 3500-5500 \AA~at a spectral resolution of 4.4 \AA~in the blue channel. 

The spectra were processed with the MIDAS package involving standard procedures (e.g., bias-subtraction, flat-fielding, wavelength calibration using He-Ar or Hg-Kr lamp spectra, flux calibration and atmospherical correction for broad molecular bands). The optimal extraction algorithm of Horne (\cite{Hor86}) was used to
extract the one-dimensional spectra. A relative flux calibration was performed using secondary standards for spectrophotometry from Oke and Gunn (\cite{Oke83}). Several spectra show residuals from night sky line subtraction (e.g., 15A and 33A). In some cases the atmospheric band correction was only partially successful (see 131Z).

The optical images ($R$ or $V$) of the new UDS objects and their optical spectra are presented in Fig. \ref{spec1}.

\section{The optical identification of UDS X-ray sources}


To identify the new 46 X-ray sources from the UDS not contained in the previously optically identified RDS sample, we apply the identification scheme described in Paper II. The scheme involves identification (ID) 
classes that characterize the spectroscopic information as 
    detailed below.

The existence of broad UV/optical emission lines with FWHM$>$1500 km s$^{-1}$ 
in the optical spectra of 27 X-ray sources of the 46 new UDS sources reveals broad emission line AGNs (type I), which are subclassified in the ID classes {\it a-c}. 
The 15 new objects of ID class {\it a} show at least two of the high-redshifted broad emission lines: Mg~II~$\lambda2798$, C~III]~$\lambda1908$, C~IV~$\lambda1548$, Si~IV~$\lambda1397$, Ly${\alpha}$~$\lambda1216$. 
Object 817A, a quasar at $z=4.45$ (Schneider et al. \cite{Schn98}), is the most distant X-ray selected quasar found to date, and object 832A is a broad absorption line quasar (BALQSO) at $z=2.735$.

Ten new objects belong to the ID class {\it b}. Their optical spectra show only a broad Mg~II emission line and usually several narrow emission lines (e.g., [Ne~V]~$\lambda3426$ and [O~II]~$\lambda3727$). The broad Mg~II lines of 513 (34O) and of 426A are marginally seen in the low S/N spectra (see Fig. \ref{spec1}), but the Gaussian fit to the lines reveals 3$\sigma$ detections in both cases. In addition, source 513 (34O) has been detected at 6 cm wavelength with the VLA (Ciliegi, private communication) confirming its AGN nature. 

\begin{figure}[htb]
     \begin{minipage}{86mm}
\psfig{figure=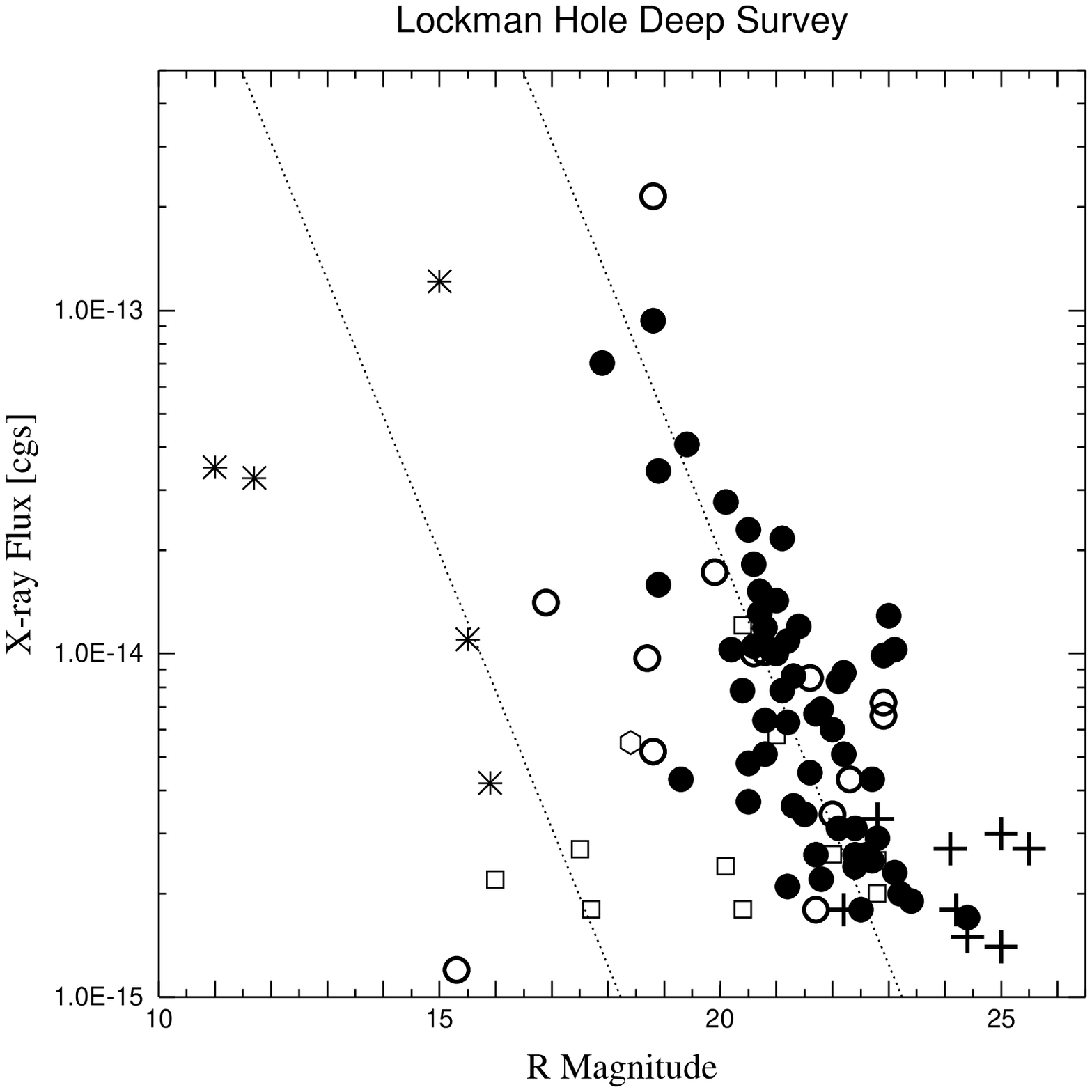,bbllx=53pt,bblly=260pt,bburx=525pt,bbury=705pt,width=85mm,clip=}
     \end{minipage}
     \caption[]{X-ray flux in the 0.5-2.0 keV band versus optical $R$ magnitudes for all objects of the UDS sample. 
The filled circles show the AGN type-I objects, open circles are AGN type-II objects. Open squares are groups and cluster of galaxies. The only galaxy in our sample is given by the hexagon. Asterisks are stars. The crosses mark the spectroscopically unidentified sources.
The dotted lines give the typical f$_{\rm x}$/f$_{\rm opt}$ ratios for stars (left) and
for AGNs (right) defined by Stocke et al. (\cite{Sto91}). }
     \label{x_rdis}
\end{figure}

There is no ID class {\it c} object (showing only broad Balmer emission lines) among the 46 new sources.       

Four X-ray sources are identified with narrow emission line AGNs (type II). One of these AGNs (901A) is identified with ID class {\it d}, where 
high ionization [Ne V]~$\lambda3426$ emission lines indicate an AGN.

Three narrow line AGNs belong to the ID class {\it e}, which contains 13 new UDS sources. The optical spectra of these sources show neither broad emission lines nor high ionization [Ne~V] emission lines. 

The identification of the ID class {\it e} objects is based  on the ratio of X-ray to optical flux $f_{\rm x}/f_{\rm v}$ as defined by Stocke et al. (\cite{Sto91}), the angular extent of the X-ray source and its high 0.5-2.0 keV X-ray luminosity (L$_{\rm x} \ge 10^{43}$ erg s$^{-1}$) indicating an AGN or a group/cluster of galaxies. 

%
%
   \begin{table*}[t]
     \caption{Photometric and spectroscopic properties of optical counterparts}
      \label{id_tab}
      \begin{flushleft}
      \begin{tabular}{l@{\hspace*{3mm}}lcc@{\hspace*{2mm}}c@{\hspace*{2mm}}c@{\hspace*{2mm}}c@{\hspace*{2mm}}c@{\hspace*{2mm}}c@{\hspace*{2mm}}c@{\hspace*{2mm}}c@{\hspace*{2mm}}c@{\hspace*{2mm}}c@{\hspace*{2mm}}c}
      \hline\noalign{\smallskip}
 name & \hspace{0.25cm}$R$ & $\alpha_{2000}$ & $\delta_{2000}$ & $\triangle$ pos & $S_{\rm x}$ & log f$_{\rm x}$/f$_{\rm v}$ & G & $z$  &  $M_{\rm v}$  &  log L$_{\rm x}$  & $R-K^{\prime}$ & class & ID class\\ \noalign{\smallskip}
\hspace*{0.2cm}(1) & \hspace{0.2cm}(2) & (3) & (4) & (5) &  (6) &  (7) & (8) & (9) & (10) & (11) & (12) & (13) & (14) \\ \noalign{\smallskip}
      \hline\noalign{\smallskip}
      \noalign{\smallskip}
2A          & 20.6$^{\ast}$  & 10 52 30.1 & 57 39 13.4 & 2H & 1.05 & -0.03 &s &1.437 & -23.8 & 44.14  & -   & AGN I&a\\
5A$^{1)}$   & 21.0$^{\ast}$ & 10 53 02.5 & 57 37 57.0 & 2H & 1.00 &  0.11 &s&1.881 & -24.0 & 44.38    & 2.7 & AGN I&a \\
6A          & 18.8  & 10 53 16.9 & 57 35 52.3 & 1H & 9.32 &  0.20 &s& 1.204 & -25.2 & 44.92  & 2.2 & AGN I&a\\
8B          & 15.0  & 10 51 31.1 & 57 34 39.4 & 1H & 12.14& -1.21 &s&       &       &        & 4.4 & STAR &M5 V\\
9A          & 21.2  & 10 51 54.5 & 57 34 37.7 & 1H & 1.09 &  0.23 &s& 0.877 & -22.2 & 43.68  & 2.9 & AGN I&b\\
\noalign{\smallskip}
11A         & 23.0:$^{\ast}$ & 10 51 08.4 & 57 33 45.4 & 0H & 1.29 &  1.02 &s& 1.540 & -21.5 & 44.30  & -   & AGN I&a\\
12A         & 22.9  & 10 51 48.8 & 57 32 48.4 & 2H & 0.72 &  0.72 &s& 0.990 & -20.7 & 43.62  & 4.9 & AGN II&d\\
13A$^{1)}$  & 22.0 & 10 52 13.3 & 57 32 21.7 & 0H & 0.60 &  0.29 &s& 1.873 & -23.0 & 44.16   & 2.5 & AGN I&a \\
14Z         & 25.0  & 10 52 42.4 & 57 31 58.4 & 1H & 0.30 &  1.19 & g:&1.94$_{\rm phot.}$ & -22.9  &  43.89  & 5.4 & -$^{2)}$     &e:\\
15A$^{1)}$  & 21.5 & 10 53 00.0 & 57 31 55.2 & 0H & 0.34 & -0.16 &s& 1.447 & -22.9 & 43.66   & 3.7 & AGN I&a \\ 
\noalign{\smallskip}
16A         & 19.4  & 10 53 39.8 & 57 31 03.9 & 2H & 4.07 &  0.08 &s& 0.586 & -23.1 & 43.87  & 3.1 & AGN I&c\\
17A         & 21.8$^{\ast}$ & 10 51 04.0 & 57 30 54.0 & 2H & 0.69 &  0.27 &s& 2.742 & -23.9 & 44.59  & -   & AGN I&a\\
18Z$^{1)}$  & 22.6 & 10 52 28.4 & 57 31 04.4 & 0H & 0.26 & 0.16  &s& 0.931 & -20.9 & 43.12   & 3.3 &  AGN I&b\\
19B         & 22.2  & 10 51 37.5 & 57 30 43.2 & 1H & 0.88 &  0.53 &s& 0.894 & -21.2 & 43.61  & 3.8 & AGN I&b\\
20C         & 15.5$^{\ast}$  & 10 54 10.4 & 57 30 37.9 & 2H & 1.10 & -2.05 & s&      &       &        & -   & STAR & M5 V\\
\noalign{\smallskip}
23A         & 22.4  & 10 52 24.7 & 57 30 09.6 & 1H & 0.26 &  0.08 &g& 1.009 & -21.3 & 43.19  & 4.4 & AGN I&b\\
24Z$^{1)}$  & 20.6$^{\ast}$ & 10 50 44.2 & 57 29 19.7 & 3P & 1.00 & -0.05  &g& 0.480 & -21.5 & 43.08   & -   & AGN II&e \\
25A         & 20.8  & 10 53 45.0 & 57 28 40.2 & 1H & 1.19 &  0.11 &s& 1.816 & -24.1 & 44.42  & 2.8 & AGN I&a\\
26A         & 18.7$^{\ast}$  & 10 50 19.8 & 57 28 12.2 & 6P & 0.97 & -0.82 &g& 0.616 & -23.9 & 43.30  & -   & AGN II&d\\
27A         & 20.8  & 10 53 50.3 & 57 27 09.2 & 0H & 0.51 & -0.26 &s& 1.720 & -24.0 & 44.00  & 2.3 & AGN I&a\\ 
\noalign{\smallskip}
28B         & 18.8$^{\ast}$  & 10 54 21.3 & 57 25 44.3 & 2H & 21.54 &  0.56 &g& 0.205 & -21.4 & 43.63 & -   & AGN II$^{3)}$&c\\
29A         & 21.1$^{\ast}$  & 10 53 35.1 & 57 25 41.6 & 1H & 2.17 &  0.49 &s& 0.784 & -22.0 & 43.88  & 2.3 & AGN I&b\\
30A         & 21.1  & 10 52 57.3 & 57 25 07.1 & 2H & 0.78 &  0.04 &s& 1.527 & -23.4 & 44.07  & 3.0 & AGN I&a\\
31A         & 20.5  & 10 53 31.8 & 57 24 53.8 & 0H & 2.30 &  0.27 &s& 1.956 & -24.5 & 44.78  & 2.8 & AGN I&a\\
32A         & 17.9  & 10 52 39.6 & 57 24 31.7 & 1H & 7.02 & -0.28 &s& 1.113 & -26.0 & 44.72  & 1.8 & AGN I&a\\
\noalign{\smallskip}
33A$^{1)}$  & 21.8 & 10 52 00.0 & 57 24 26.1 & 2H & 0.22 & -0.23 &g& 0.974 & -21.8 & 43.09   & 3.8   & AGN I&b \\
34F$^{1)}$  & 20.1 & 10 52 55.3 & 57 24 09.7 & 29H & 0.24 & -0.87 &g& 0.262 & -20.7   & 41.90  & 2.4 & GRP&e \\
35A         & 18.9$^{\ast}$  & 10 50 39.6 & 57 23 36.3 & 1R & 1.59 & -0.53 &s& 1.439 & -25.5 & 44.32  & -   & AGN I&a\\
37A         & 20.1  & 10 52 48.2 & 57 21 17.4 & 3H & 2.77 & 0.19  &s& 0.467 & -21.9 & 43.49  & 2.8 & AGN I&b\\
38A         & 21.3  & 10 53 29.5 & 57 21 03.9 & 2H & 0.86 & 0.16 &s& 1.145 & -22.6 & 43.84   & -   & AGN I&a\\
\noalign{\smallskip}
39B$^{1)}$  & 21.7 & 10 52 09.8 & 57 21 04.7 & 2H & 0.26 &-0.20 &s& 3.279 & -24.4 & 44.34    & 1.1 & AGN I&a \\
41C         & 17.5  & 10 53 18.7 & 57 20 43.9 & 5H & 0.27 & -1.86 &g& 0.340 & -23.8 & 42.19  & -   & GRP&e\\
42Y         & 20.7$^{\ast}$  & 10 50 16.1 & 57 19 53.8 & 8P & 1.52 &  0.17 &s& 1.144 & -23.2 & 44.08  & -   & AGN I&a\\ 
43A         & 22.9  & 10 51 05.1 & 57 19 23.2 & 1R & 0.99 &  0.87 &s& 1.750 & -21.9 & 44.31  & -   & AGN I&a\\
45Z         & 21.2  & 10 53 18.9 & 57 18 50.0 & 2H & 0.63 & -0.01 &g& 0.711 & -21.7 & 43.25  & -   & AGN II&d\\
\noalign{\smallskip}
46A         & 23.1  & 10 51 20.1 & 57 18 47.9 & 2R & 1.03 &  0.96 &s& 1.640 & -21.6 & 44.26  & -   & AGN I&a\\
47A         & 22.2  & 10 52 45.0 & 57 17 33.4 & 5H & 0.51 &  0.30 &s& 1.058 & -21.6 & 43.53  & -   & AGN I&a\\
48B         & 19.9$^{\ast}$  & 10 50 46.2 & 57 17 33.1 & 1R & 1.73 & -0.09 &g& 0.498 & -22.2 & 43.35  & -   & AGN II&e:\\ 
51L         & 21.6$^{\ast}$  & 10 51 17.0 & 57 15 51.4 & 3P & 0.85 &  0.28 &s& 0.620 & -21.0 & 43.25  & -   & AGN II&d\\
52A         & 21.0  & 10 52 43.3 & 57 15 44.6 & 2H & 1.43 & 0.27  &s& 2.144 & -24.2 & 44.66  & -   & AGN I&a\\
\noalign{\smallskip}
53A         & 18.4$^{\ast}$  & 10 52 06.3 & 57 15 24.7 & 5P & 0.55 & -1.19 &g& 0.245 & -22.2 & 42.20  & -   & GAL&e\\
54A         & 21.4$^{\ast}$  & 10 53 07.4 & 57 15 04.6 & 2H & 1.20 &  0.35 &s& 2.416 & -24.1 & 44.71  & -   & AGN I&a\\
55C         & 20.9$^{\ast}$  & 10 50 09.4 & 57 14 43.3 & 1P & 1.03 & 0.08  &s:& 1.643 & -23.8 & 44.26 & -   & AGN I& a\\ 
56D         & 18.9$^{\ast}$  & 10 50 20.2 & 57 14 21.7 & 1R & 3.41 & -0.20 &s& 0.366 & -22.6 & 43.36  & -   & AGN I&b\\ 
58B         & 21.0  & 10 52 38.8 & 57 12 59.7 & 10P& 0.58 & -0.13  &g& 0.629 & -21.6 & 43.09  & -   & GRP&e\\
\noalign{\smallskip}
59A         & 16.9$^{\ast}$  & 10 53 24.8 & 57 12 30.7 & 5P & 1.41 & -1.38 &g& 0.080 & -21.3 & 41.61  & -   & AGN II$^{3)}$&c\\
60B        & 21.7$^{\ast}$ & 10 52 48.5 & 57 12 06.0 & 3P & 0.67 &  0.22 &s& 1.875 & -23.3 & 44.20   & -   & AGN I&a\\
61B        & 20.8$^{\ast}$ & 10 51 26.3 & 57 11 31.1 & 6P & 0.64 & -0.16 &s& 0.592 & -21.7 & 43.08   & -   & AGN I&b\\ 
62A        & 11.0$^{\ast}$ & 10 52 01.3 & 57 10 45.9 & 2R & 3.49 & -3.35 &s&       &       &         & -   & STAR & K0 V\\
67B(s)$^{\ast\ast}$     & 20.5:$^{\ast)}$& 10 50 54.3 & 57 06 51.5 & 8P & 1.21 & -0.01 &g&0.550 & -21.9 & 43.29 & -   & GRP&e\\
      \noalign{\smallskip}
      \hline
      \end{tabular}
      \end{flushleft}
$^{\rm 1)}$ New UDS sources, $^{\rm 2)}$ Photometric redshift determination indicates an AGN type II,\\ $^{3)}$ AGN type II because of its large Balmer decrement (see text).\\
$^{\ast)}$ $R$ magnitudes not based on Keck photometry.
$^{\ast\ast)}$ Paper II gives the coordinates of 67A instead of 67B(s).
   \end{table*}
\addtocounter{table}{-1}

%
%
   \begin{table*}[t]
     \caption{(continued).}
      \begin{flushleft}
     \begin{tabular}{l@{\hspace*{3mm}}lcc@{\hspace*{2mm}}c@{\hspace*{2mm}}c@{\hspace*{2mm}}c@{\hspace*{2mm}}c@{\hspace*{2mm}}c@{\hspace*{2mm}}c@{\hspace*{2mm}}c@{\hspace*{2mm}}c@{\hspace*{2mm}}c@{\hspace*{2mm}}c}
      \hline\noalign{\smallskip}
 name & \hspace{0.25cm}$R$ & $\alpha_{2000}$ & $\delta_{2000}$ & $\triangle$ pos & $S_{\rm x}$ & log f$_{\rm x}$/f$_{\rm v}$ & G & $z$  &  $M_{\rm v}$  &  log L$_{\rm x}$  & $R-K^{\prime}$ & class & ID class\\ \noalign{\smallskip}
\hspace*{0.2cm}(1) &\hspace{0.2cm} (2) & (3) & (4) & (5) &  (6) &  (7) & (8) & (9) & (10) & (11) & (12) & (13) & (14)\\ \noalign{\smallskip}
      \hline\noalign{\smallskip}
      \noalign{\smallskip}
70A$^{1)}$       & 20.6$^{\ast}$ & 10 52 15.7 & 57 04 03.6 & 7P & 1.82 & 0.21  &s&1.010 & -23.1 & 44.04 & -   & AGN I&b\\
73C              & 20.7 & 10 50 09.6 & 57 31 43.5 & 14P& 1.31 & 0.11  &s:&1.561 & -23.9 & 44.32& -   & AGN I&a\\
75A$^{1)}$       & 19.3 & 10 51 25.4 & 57 30 50.7 & 1H & 0.43 & -0.94 &s&3.409 & -26.9 & 44.60 & 1.7 & AGN I&a \\
77A              & 22.1 & 10 52 59.3 & 57 30 30.2 & 1H & 0.83 &  0.47 &s&1.676 & -22.6 & 44.19 & 3.3 & AGN I&a\\ 
80A$^{1)}$       & 21.2 & 10 51 44.7 & 57 28 06.6 & 1H & 0.21 & -0.49 &s&3.409 & -25.0 & 44.28 & 2.3   & AGN I&a \\
\noalign{\smallskip}
82A$^{1)}$       & 22.8 & 10 53 12.3 & 57 25 06.4 & 1H & 0.29 &  0.29 &s&0.960 & -20.8 & 43.19 & 3.8 & AGN I&b \\
84Z              & $>$25.5&10 52 17.1& 57 20 16.8 & 1H & 0.27 &$>$1.34&g:&2.71$_{\rm phot.}$& -24.0 & 44.17 & 6.2 &      -$^{2)}$&e\\
104A$^{1)} $     & 18.8$^{\ast}$ & 10 52 41.6 & 57 36 50.2 & 2H & 0.52 & -1.05 &g&0.137 & -20.6 & 41.65 & -   & AGN II&e \\
117Q             & 22.9 & 10 53 48.8 & 57 30 33.9 & 1H & 0.66 & 0.69  &s&0.780 & -20.2 & 43.35 & 4.9 & AGN II&d\\
120A$^{1)}$      & 20.4 & 10 53 09.4 & 57 28 20.9 & 1H & 0.78 & -0.24 &s&1.568 & -24.2 & 44.10 & 2.7 & AGN I&a  \\
\noalign{\smallskip}
128E$^{1)}$      & 22.0: & 10 53 50.8 & 57 25 12.7 & 3H & 0.26 & -0.08 &g&0.478 & -20.1 & 42.49 & -   & GRP:&e:\\
131Z$^{1)}$      & 17.7 & 10 53 40.0 & 57 23 52.8 & 5H & 0.18 & -1.95 &g&0.205 & -22.5 & 41.55 & -   & CLUS&e \\
151B$^{1)}$      & 20.2:$^{\ast}$& 10 54 06.4 & 57 13 40.4 & 3P & 1.03 & -0.20 &s&1.200 & -23.8 & 43.96 & -   & AGN I&b \\ 
228Z$^{1)}$      & 22.8 & 10 53 40.2 & 57 35 18.2 & 8H & 0.25$^{5)}$ &0.23 &g&$\sim$1.263& -21.3 &  43.39 & 5.2 & CLUS&e\\
229Z$^{1)}$      & 22.8 & 10 53 46.7 & 57 35 15.5 & 2H & 0.20$^{5)}$ & 0.13  &g&$\sim$1.263& -21.3 & 43.30 & 5.6 & CLUS&e \\
\noalign{\smallskip}
232A$^{1)}$ & 11.7      & 10 53 36.6 & 57 38 00.8 & 2H & 3.25 & -3.10 &   s  & &                      & & 2.4 & STAR & F7 III \\
426A$^{1)}$      & 22.0$^{\ast}$ & 10 53 03.8 & 57 29 24.6 & 1H & 0.34 & 0.04  &g&0.788 & -21.1 & 43.08 & 4.5 & AGN I&b \\
428E$^{1)}$      & 22.4 & 10 53 24.7 & 57 28 18.3 & 1H & 0.31 &  0.16 &s&1.518 & -22.1 & 43.66 & 3.6 & AGN I&a\\ 
434B$^{1)}$      & 22.2 & 10 52 58.3 & 57 22 51.1 & 2H & 0.18 & -0.15 &s&-          &      -&-      & 4.1 &      -&e\\
477A$^{1)}$      & 20.5$^{\ast}$ & 10 53 06.2 & 57 34 24.7 & 1H & 0.37 & -0.52 &s&2.949 & -25.4 & 44.39 & 1.9 & AGN I&a \\
\noalign{\smallskip}
486A$^{1)}$      & 24.4 & 10 52 43.5 & 57 28 00.0 & 1H & 0.15 & 0.65  &s&1.21$_{\rm phot.}$ & -22.4   & 43.13  & 5.3 & -$^{2)}$&e \\
504~(51D)        & 20.8$^{\ast}$ & 10 51 14.5 & 57 16 15.5 & 1R & 1.01 &  0.03 &s&0.528 & -21.5 & 43.17 & -   & AGN II&d\\
513$^{1)}$~(34O) & 22.3 & 10 52 54.4 & 57 23 42.2 & 1H & 0.43 & 0.26  &s&0.761 & -20.8 & 43.14 & 4.1 & AGN I&b\\ 
607$^{1)}$~(36Z) & 24.1 & 10 52 20.1 & 57 23 06.8 & 1H & 0.27 &  0.78 &s&-&-& -& 4.3 & -$^{4)}$  &  e\\
634A$^{1)}$      & 23.2 & 10 53 11.8 & 57 23 06.2 & 1H & 0.20 & 0.29  &s&1.544 & -21.4 & 43.49 & -   & AGN I&b\\
\noalign{\smallskip}
801A$^{1)}$      & 22.1 & 10 52 45.4 & 57 37 45.7 & 3H & 0.31 &  0.04 &s:&1.677 & -22.6 & 43.76& -   & AGN I&a\\
802A$^{1)}$      & 15.9 & 10 52 22.4 & 57 37 35.4 & 1H & 0.42 & -2.31 &s&      &       &       & -   &STAR & M0 V:\\
804A$^{1)}$      & 22.7$^{\ast}$ & 10 53 12.5 & 57 34 24.4 & 0H & 0.43 & 0.42 &s:&1.213 & -21.3 & 43.59& 3.8 & AGN I&b\\
805A$^{1)}$      & 24.4 & 10 53 47.8 & 57 33 54.7 & 3H & 0.17  & 0.70  &s&2.586 & -21.2  & 43.92  & 4.7 & AGN I&a\\
814~(37G)        & 20.5 & 10 52 44.8 & 57 21 23.2 & 1H & 0.48 & -0.41 &s&2.832 & -25.3 & 44.46 & 1.5 & AGN I&a\\
\noalign{\smallskip}
815C$^{1)}$      & 20.4 & 10 53 29.1 & 57 35 36.0 & 4H & 0.18 & -0.87 &g&0.700 & -22.5 & 42.69 & 4.0 & CLUS&e \\
817A$^{1)}$      & 23.1 & 10 52 25.9 & 57 19 06.7 & 2H & 0.23 &  0.31 &s&4.450 & -23.6 & 44.58 & 3.9 & AGN I&a \\
821A$^{1)}$      & 22.7 & 10 53 22.3 & 57 28 51.5 & 1H & 0.25 &  0.19 &s&2.300 & -22.7 & 43.98 & 4.0 & AGN I&a \\ 
825A$^{1)}$      & 22.8 & 10 52 06.5 & 57 22 14.3 & 2H & 0.33 &  0.35 &s& -    &   -   &   -   & 4.4   & - &e\\ 
827B$^{1)}$      & 21.7$^{\ast}$ & 10 53 03.9 & 57 35 31.1 & 2H & 0.18 & -0.35  &g&0.249 & -18.9 & 41.73 & 4.1 & AGN II &e\\
\noalign{\smallskip}
828A$^{1)}$      & 21.6 & 10 53 57.3 & 57 32 42.1 & 2H & 0.45 & 0.00  &s&1.282 & -22.6 & 43.66 & -   & AGN I&a \\
832A$^{1)}$      & 22.4$^{\ast}$ & 10 52 07.5 & 57 38 38.1 & 4H & 0.24 & 0.05  &s&2.730 & -23.3 & 44.13 & -   & AGN I&a \\ 
837A$^{1)}$      & 23.4 & 10 51 37.5 & 57 28 56.4 & 1H & 0.19 & 0.35  &s&2.013 & -21.7 & 43.73 & 3.7 & AGN I&a\\ 
840D$^{1)}$      & 16.0:$^{\ast}$& 10 52 46.9 & 57 40 42.0 & 7H & 0.22 & -2.55 &g&0.074 & -22.0 & 40.73 & -   & GRP:&e\\ 
861A$^{1)}$      & 22.5 & 10 53 58.4 & 57 29 25.8 & 1H & 0.18 &-0.03  &s&1.843 & -22.4 & 43.62 & -   & AGN I&a\\
\noalign{\smallskip}
866A$^{1)}$ & 24.2$^{\ast}$ &  10 52 36.7 &  57 37 47.6  & 2H &  0.36 &  0.95  &s&  -   &  -    &  -    &  -  &   -  & e\\
870$^{1)}$~(36F) & 21.3 & 10 52 25.3 & 57 22 47.1 & 2H & 0.36 & -0.21 &g&0.807 & -21.9 & 43.12 & 3.6 & AGN I&b \\ 
901A$^{1)}$      & 15.3 & 10 52 52.8 & 57 29 00.3 & 1H & 0.12 & -3.09 &g&0.204 & -24.9 & 41.37 & 3.4 & AGN II&d \\ 
905A$^{1)}$      & 25.0 & 10 52 51.4  & 57 20 11.4  & 2H   &  0.14  & 0.86 &s&  -   &  -    &   -   & 6.3 &   -  &e \\
      \noalign{\smallskip}
      \hline
      \end{tabular}
      \end{flushleft}
$^{\rm 4)}$ See text in Sect. 6, $^{\rm 5)}$ PSPC flux is larger than this HRI flux value (see Table \ref{xray_tab}). 
   \end{table*}
\clearpage

Seven of the new class e objects are spectroscopically confirmed as groups or clusters of galaxies (see notes on individual sources below). 
In two cases we found stars as optical counterparts of the X-ray sources (see Table \ref{id_tab}).

Six new UDS sources (434B, 486A, 607, 828A, 866A, and 905A ), which belong to class e, still lack an optical identification.

Fig. \ref{x_rdis} shows the 0.5-2.0 keV X-ray flux of all UDS sources as a function of the $R$ magnitude of their optical counterparts. 
Nearly all AGNs are located around a well defined line with $f_{\rm x}/f_{\rm v}=1$. 

Six of the eight unidentified UDS sources have $R$ magnitudes lower than the spectroscopic limit of our survey, which is approximately $R\sim 23.5$. The photometric redshift
determination of these sources are presented in Sect. 6.
The identification of several new UDS sources is discussed in more detail below.

\subsection{Notes on individual objects}

{\bf Source 24} is detected as a point source with the PSPC and is not covered by the HRI exposure.  The Palomar $V$ image shows one object (24Z) inside the PSPC error circle. The optical spectrum contains narrow emission lines of Oxygen and a H$\beta$ emission line at $z=0.480$. The Gaussian fit to the H$\beta$ line reveals a significant broader line (FWHM$=$1440 km~s$^{-1}$) compared to the width of the Oxygen emission lines (see Table \ref{FWHM_opt}). The log $f_{\rm x}/f_{\rm v}$ ratio and the X-ray luminosity of  L$_{\rm x}>10^{43}$ erg~s$^{-1}$ in the 0.5-2.0 keV band are rather high for a normal galaxy.
We classify the object as an AGN. 

{\bf Source 34} is extended in the HRI exposure. It was not possible to determine the morphology with the PSPC, because of the confusion with source 513 at a distance of about 35$''$. There is no optical or even $K^{\prime}$ band counterpart inside the HRI error circle. Two galaxies 34L and 34M are located at distances of $\sim5''$ and $\sim18''$ from the HRI position. The object 34M is a narrow emission line galaxy at $z=0.263$. A 1 hour long slit exposure of the object 34L has not yielded a definitive redshift. Another emission line galaxy (34F) at $z=0.262$ is at a distance of $\sim28''$ from the HRI position. The ratio log $f_{\rm x}/f_{\rm v}=-0.87$ for 34F is consistent with those of groups or clusters of galaxies. We tentatively classify source 34 as a group of galaxies.  
 
{\bf Source 104} is consistent with a point source in the PSPC and the HRI exposures. The relatively bright galaxy 104A is located inside the HRI circle and shows several narrow emission lines at $z=0.137$. 
The flux ratios of the emission lines log ([O~III]~$\lambda$5007/H$\beta~\lambda4861)=0.86$, log ([N~II]~$\lambda6584$/H$\alpha~\lambda6563)=-0.24$ and log ([S~II]~$\lambda6716+\lambda6731$/~H$\alpha~\lambda6563)=-0.50$ reveal an AGN using the diagnostic diagrams by Veilleux \& Osterbrock (\cite{Vei87}).
The ratio log $f_{\rm x}/f_{\rm v}=-1.05$ is at the lower limit of the EMSS AGN (Stocke et al. \cite{Sto91}). This object has a 0.5-2.0 keV X-ray luminosity of log L$_{\rm x}=41.65$ (erg~s$^{-1}$) and is one of the four low luminosity AGN found in the survey (see Fig. \ref{logLxMv}). A second galaxy (104C) at $z=0.134$ is at a distance of 25$''$ from the X-ray source. Although the AGN is probably the source of the X-ray radiation, object 104A could belong to a group of galaxies. 

{\bf Source 128} appears significantly extended in the HRI image, but it is not detected in the PSPC exposure. However, we are not able to exclude that the extended emission results from source confusion. 
The galaxy 128E inside the HRI error circle shows a narrow [O~II]~$\lambda3727$ emission line at redshift $z=0.478$. 
The ratio log $f_{\rm x}/f_{\rm v}=-0.08$ agrees with the range of galaxy groups and clusters of galaxies. The X-ray extent of the source cannot originate from the galaxy 128E at such a high redshift. 

A second galaxy (128D) at a redshift $z=0.031$ is located near the edge of the HRI error circle. The spectrum of the object shows several narrow emission lines of Oxygen and the Balmer series. The flux ratios  log ([O~III]~$\lambda$5007/H$\beta~\lambda4861)=0.46$ and log ([S~II]~$\lambda6716+\lambda6731$/H$\alpha~\lambda6563)=-1.13$ allow a secure classification as a starburst galaxy using the Veilleux \& Osterbrock diagnostic diagrams (1987). The galaxy 128D is probably a field object.

There are three additional faint galaxies within a distance of 20$''$ from the galaxy 128E, which seems to indicate the presence of a poor group of galaxies with an X-ray luminosity of 3.1 $\times$ 10$^{42}$ (see, e.g, Mulchaey et al. \cite{Mul96}) or a fossil galaxy group similar to that object found by Ponman et al. (\cite{Pon94}). We tentatively classify source 128 as a galaxy group.
  
{\bf Source 131} shows extended X-ray emission in the HRI exposure. The relatively bright elliptical galaxy 131Z ($z=0.205$) is located at the edge of the HRI error circle and is surrounded by at least 10 fainter galaxies. A similar object is the elliptical cD galaxy in A2218 at $z=0.175$ (Le Borgne et al. \cite{LeB92}). We classify source 131 as a cluster of galaxies. However, the ratio log $f_{\rm x}/f_{\rm v}=-1.95$ is rather low for a galaxy cluster. 

{\bf Sources 228/229}, the most extended X-ray sources of our sample, are separated only by about 1$'$. The Keck $R$-image shows a marginal excess of galaxies brighter than $R=24.5$, which indicates a high-redshift cluster of galaxies. The optical spectrum of the brightest galaxy close to the center of source 229 (left peak in Fig.1 of Paper IV) revealed a gravitationally lensed arc at $z=2.570$ (Hasinger et al. \cite{Has99b}, Paper IV). Near-infrared images have shown a number of relatively bright galaxies with $R-K^{\prime}=5.5-5.7$ (Lehmann et al. \cite{Leh00b}). 

Recently, Thompson et al. (\cite{Tho01}) have detected a broad H$\alpha$ emission line at $z=1.263$ in the near-infrared spectrum of one of the galaxies. 
The X-ray and optical morphologies suggest the presence of either a bimodal cluster or two clusters in interaction. This would be one of the two most distant X-ray selected cluster of galaxies found to date. 
Another cluster at $z=1.267$ was discovered in the ROSAT Deep Cluster Survey (Rosati et al. \cite{Ro99}).

{\bf Source 815} appears pointlike in the HRI exposure. The optically brightest galaxy 815C inside the HRI error circle is an elliptical galaxy at $z=0.700$. Two further narrow emission line galaxies (815D and 815F) have been found at the same redshift. The ratio log $f_{\rm x}/f_{\rm v}=-0.87$ for 815C is consistent with a cluster of galaxies or an AGN. The galaxy 815C has been detected in the VLA 6cm survey of the Lockman Hole (Ciliegi, private communication), which reveals a radio galaxy. At least 7 additional faint galaxies are within a distance of 20$''$ from the center of X-ray position, which seems to indicate a rich galaxy cluster around the radio galaxy.  We cannot conclude that the detected emission is due to a single AGN or due to a cluster of galaxies. Extended X-ray emission is probably not detected because of the faintness of this source. On the base of the existense of three galaxies at $z=0.700$ we classify source 815 as a cluster of galaxies.

{\bf Source 827} is consistent with a point source in the HRI exposure. The HRI error circle contains a galaxy (827B) at $z=0.249$. The optical spectrum of 827B shows only a very strong H$\alpha$ emission line, but no further emission lines resulting in a large Balmer decrement. The H$\alpha$ line width of ~1200 km~s$^{-1}$ (FWHM) is significantly broader than typical narrow emission lines, which indicates an AGN.  The source is detected in the VLA 20 cm survey of the Lockman Hole region (DeRuiter et al. \cite{deRui97}). The ratio log $f_{\rm x}/f_{\rm v}=-0.35$ of 827B agrees with the range of AGN. We classify source 827 as an AGN. It is one of the four low-luminosity AGN (log L$_{\rm x} \le 42$ erg~s$^{-1}$) of our sample (see Fig. \ref{logLxMv}).
  
{\bf Source 840} is extended in the HRI pointing and not covered by the inner PSPC field of view. The HRI error circle contains a bright star and a faint galaxy (840C) at $z=0.074$. A brighter galaxy (840D) at the same redshift is located at a distance of 20$''$ from the center of the HRI position. The ratio log $f_{\rm x}/f_{\rm v}=-2.55$ for 840D and the X-ray luminosity (log $L_{\rm x}=40.7$) are very low for a group of galaxies.   
Due to the lower sensitivity of the HRI to detect extended X-ray emission the X-ray flux could be underestimated by a factor of 2-8 (see, e.g., sources 41 and 228 in Table \ref{xray_tab}). 
Source 840 is classified as a group of galaxies due to the extended X-ray emission. However, we cannot exclude that the X-ray emission is due to a single galaxy
 or a to the sum of the emission of a few galaxies in the group. 

%
%
   \begin{table*}[t]
      \caption{Emission line properties EW in [\AA] and FWHM in [km~s$^{-1}$] of RDS and UDS AGNs.}
      \label{line_comp}
      \begin{flushleft}
      \begin{tabular}{llrrrrrrrr}
      \hline\noalign{\smallskip}
 &  & \multicolumn{4}{c}{RDS AGNs} &  \multicolumn{4}{c}{UDS AGNs}\\ \noalign{\smallskip}
 \noalign{\smallskip}
\hspace{0.4cm}lines  & component  & \multicolumn{2}{c}{FWHM} & \multicolumn{2}{c}{EW} & \multicolumn{2}{c}{FWHM} & \multicolumn{2}{c}{EW}\\
        &   & mean\hspace{0.4cm} & n\hspace{0.12cm} & mean\hspace{0.1cm} & n\hspace{0.12cm} &mean\hspace{0.4cm} & n\hspace{0.12cm} &mean\hspace{0.1cm} & n\hspace{0.2cm} \\
\hspace{0.5cm}(1) &\hspace{0.4cm} (2) & (3)\hspace{0.5cm} & \hspace{0.1cm} (4) & (5)\hspace{0.2cm} & (6) & (7)\hspace{0.5cm} & (8) & (9)\hspace{0.2cm} & (10)\\ \noalign{\smallskip}
\hline \noalign{\smallskip}
Ly$\alpha~\lambda1216$ &broad&     -      & 2&    -    & 2&2600$\pm$230& 8&35$\pm$7&8\\
Si~IV~$\lambda1397$    &broad&4780$\pm$450& 4&12$\pm$2 & 4&4700$\pm$390&10&16$\pm$4&10\\
C~IV~$\lambda1548$     &broad&4030$\pm$530& 9&37$\pm$6 & 9&3980$\pm$360&23&32$\pm$4&23\\
C~III$]~\lambda1908$   &broad&5380$\pm$390&16&41$\pm$15&16&4820$\pm$340&27&31$\pm$9&27 \\
Mg~II~$\lambda2798$    &broad&4190$\pm$370&21&46$\pm$12&21&4230$\pm$300&34&39$\pm$8&34 \\ \noalign{\smallskip}
$[$Ne~V$]~\lambda3426$ &narrow& 680$\pm$ 90& 8& 4$\pm$1 &9 & 680$\pm$ 90& 8& 5$\pm$1&14 \\
$[$O~II$]~\lambda3727$ &narrow& 440$\pm$ 60&11&12$\pm$3 &17& 410$\pm$ 40&16&12$\pm$2&25 \\
$[$Ne~III$]~\lambda3868$&narrow&820$\pm$180& 5& 5$\pm$2 &8 & 820$\pm$180& 5& 6$\pm$1& 11 \\
H$\gamma~\lambda4340$  &narrow&     -      & 2& 4$\pm$1 &5 &      -     & 2& 4$\pm$1&6\\
H$\beta~\lambda4861$   &narrow&     -      & 1& 6$\pm$2 &4 &      -     & 2& 6$\pm$1&5\\ \noalign{\smallskip}
                       &broad&3810$\pm$1300&4&38$\pm$9 &4 &3810$\pm$1300&4&38$\pm$9&4\\ 
$[$O~III$]~\lambda4959$&narrow& 380$\pm$120& 5& 6$\pm$2 & 8& 420$\pm$100& 6& 6$\pm$2&10\\
$[$O~III$]~\lambda5007$&narrow& 310$\pm$ 30& 5&20$\pm$9 & 9& 310$\pm$ 30& 5&19$\pm$8&11\\
H$\alpha~\lambda6563$  &narrow&     -      & 0&    -    & 2&     -      & 2&74$\pm$63&3\\
                       &broad&     -      & 1&    -    & 1&     -      & 1&    -   &1\\
      \noalign{\smallskip}
      \hline
      \end{tabular}
      \end{flushleft}
      \end{table*}
%
 
{\bf Source 901} is identified with ID class {\it d}.
A high ionization [Ne V]~$\lambda3426$ emission line indicates an AGN in this source.
The flux ratios of the emission lines log~([O~III]~$\lambda$5007/H$\beta~\lambda4861)>1.31$ and log~([N~II]~$\lambda6584$/H$\alpha~\lambda6563)=0.17$ confirm the optical identification as an AGN (see diagnostic diagrams by Veilleux \& Osterbrock \cite{Vei87}).
Its X-ray luminosity of $2.3 \cdot 10^{41}$ erg~s$^{-1}$ in the 0.5-2.0 keV energy band and its ratio log $f_{\rm x}/f_{\rm v}=-3.09$ are very low for an AGN, which could be due to intrinsic absorption. The object is one of the four low luminosity AGNs (log L$_{\rm x}<42$) of the total sample of 69 UDS AGNs (see Fig. \ref{logLxMv}).\\
 
The following six new UDS X-ray sources are still spectroscopically unidentified:

{\bf Source 434} is a faint X-ray source detected in the HRI and the PSPC pointing. The HRI error circle contains two objects (434A and 434B). The optical spectrum of 434A reveals an elliptical galaxy at $z=0.762$. The ratio log $f_{\rm x}/f_{\rm v}=-0.15$ is rather high for a galaxy, but consistent with an AGN or a group or cluster of galaxies. We have no optical spectrum of 434B ($R=22.4$), which has $R-K^{\prime}=4.1$. The galaxy 434A is not detected on the $K^{\prime}$ image. We need to take an optical spectrum of 434B to identify the source 434.

{\bf Source 486} is detected as a point source in the HRI and the PSPC exposures. The $R$ images (see Fig. \ref{spec1}) show the faint object 486A ($R=23.8$) inside the HRI error circle. This object has a $K^{\prime}$ magnitude of 18.5; $R-K^{\prime}=5.3$. Source 486 is one of the three unidenitified soures with $R-K^{\prime}$ indices larger than 4.5 (14Z and 84Z being the other two). We argue that 486A is probably an obscured AGN (see photometric redshift determination in Sect. 6.). The ratio log $f_{\rm x}/f_{\rm v}=0.65$ is too high for a galaxy. 

{\bf Source 607} has been detected in the HRI and the PSPC pointing. The $R$ images show a very faint counterpart ($R=24.1$) inside the HRI error circle. Due to the high accuracy of the HRI position we argue that 36Z is the optical counterpart. We were unable to determine the spectroscopic redshift of 36Z with a one hour Keck exposure. The high ratio log $f_{\rm x}/f_{\rm v}=0.78$ is consistent with an AGN. The photometric redshift estimation of the object 607 (36Z) is described in Sect. 6. 

{\bf Source 825} is a point source in the HRI image. The HRI error circle contains a faint object 825A ($R=22.8$). We were not able to determine the redshift of 825A with a one hour Keck exposure. The high ratio log $f_{\rm x}/f_{\rm v}=0.35$ would be consistent with an AGN or with a group/cluster of galaxies. 

{\bf Source 866} is a HRI point source. The faint object 866A ($R=24.2$) is located inside the HRI error circle, but hard to see in Fig. \ref{spec1}. The high ratio log $f_{\rm x}/f_{\rm v}$=0.95 is consistent with an AGN or a group/cluster of galaxies. However, due to its X-ray faintness source 866 could be a spurious
detection. We expect about 1-2 spurious detections in our sample of 94 X-ray
spources down to a limitimg flux of $1.2 \cdot 10^{-15}$ erg~cm$^{-2}$~s$^{-1}$ in the 0.5-2.0 keV band.

{\bf Source 905} is detected as a point source in the HRI exposure. The faint
optical counterpart 905A ($R=25.0$) shows a very red colour ($R-K^{\prime}=6.3$), which could indicate an obscured AGN similar to 14Z or 84Z . The high ratio log $f_{\rm x}/f_{\rm v}$=0.86 is consistent with an AGN or a group/cluster of galaxies.

\subsection{The catalogue of optical counterparts} 

Table \ref{id_tab} contains the optical properties of the 94 X-ray sources in the UDS sample as defined in Sect. 2.1. The table is sorted by increasing internal X-ray source number (see Table \ref{xray_tab}), which is included in the name of the optical object identified with the X-ray source. 
The first four columns give the name of the object, the $R$ magnitude of the object and its right ascension and declination at epoch 2000. 
We have marked the $R$ magnitudes, which are determined from Keck LRIS images. The other values have to be regarded as somewhat uncertain. 

Column 5 gives the distance of the optical counterpart from the position of the X-ray source in arcsec. The capital letters (H, P and R) mark whether the source position is mainly derived from the 1112 ksec HRI exposures, the 207 ksec PSPC exposure or the 205 ksec HRI raster scan. The HRI or the PSPC flux in units of 10$^{-14}$ erg cm$^{-2}$ s$^{-1}$ (0.5-2.0 keV) depending if the source belongs to the HRI sample or to the PSPC samples is given in column 6. 
The next column shows the X-ray to optical flux ratio $f_{\rm x}/f_{\rm v}$ as defined by Stocke et al. (\cite{Sto91}). Columns 8 and 9 give a optical morphological parameter of the object (s$=$star-like, g$=$extended, galaxy-like) and the redshift. The morphological parameter has changed for several objects (e.g., 9A and 60B) compared to that of Paper II due to availability of higher quality Keck images.

 The absolute magnitude $M_{\rm v}$ (using the assumption $V-R=+0.22$, corresponding to a power law spectral index of -0.5) and the logarithm of the X-ray luminosity in the 0.5-2.0 keV energy band in units of erg~s$^{-1}$ (assuming an energy spectral index of -1.0) are shown in columns 10 and 11. For the three very red sources with photometric redshifts we have determined the absolute magnitude $M_{\rm v}$ using their $K$ magnitudes. The $R-K^{\prime}$ colour of the counterparts is given, when available, in column 12. Column 13 gives the optical classification (broad emission line AGNs -- AGN I, narrow emission line AGNs -- AGN II, GAL - galaxy and GRP/CLUS -- group/cluster of galaxies). 

The large Balmer decrement, indicated by the large ratio of the H$\alpha$ to the H$\beta$ emisssion line equivalent widths ($>9$), of some AGNs (e.g., 28B, 59A) reveals a large amount of optical absorption. We therefore classify these objects as AGN type II. The ID class of the optical counterparts is given in the last column. 

\subsection{AGN emission lines and galaxy absorption lines}

The optical identification of the UDS X-ray sources depends on the accurate
measurement of the optical emission line properties. Nearly all identified X-ray sources in the UDS have high S/N Keck spectra, which would allow detection of faint high ionized [Ne V] $\lambda\lambda3346/3426$ emission lines, if they were present. 

To help assign a reliable identification we have derived the emission line properties, the FWHM in km~s$^{-1}$ corrected for instrumental resolution and the rest frame EW in \AA. For the determination of the line parameters we fitted single or double Gaussian profiles using the Levenberg-Marquardt algorithm (Press et al. \cite{Pre92}), see Paper III for details. The parameter set was accepted if the reduced $\chi^{2}$ was less than 3.0. 

We distinguish between narrow and broad emission lines at FWHM of 1500 km~s$^{-1}$. In some cases we found very broad components with FWHM $>$ 8000 km~$^{-1}$. Tables \ref{FWHM_UV}-\ref{EW_opt} (see Appendix) contain the FWHM and the EW values of those emission lines, which were detected at the 3$\sigma$ level.

   \begin{table}[t]
      \caption{Galaxy absorption lines and D(4000) index.}
      \label{gal_abs}
      \begin{flushleft}
    \begin{tabular}{l@{\hspace*{3mm}}l@{\hspace*{3mm}}c@{\hspace*{3mm}}c@{\hspace*{3mm}}c@{\hspace*{3mm}}c}
      \hline\noalign{\smallskip}
name & \hspace{0.2cm}class &  $z$ &  Lines & D(4000) \\
\hspace{0.2cm}(1) & \hspace{0.3cm}(2) & (3) & (4) & (5)  \\\noalign{\smallskip}
\hline \noalign{\smallskip}
18Z & AGN I &  0.931 & $^{\ast)}$  &  - \\
24Z & AGN II&  0.480 & AB:C: & 1.43$\pm$0.26\\
33A & AGN I &  0.974 &  $^{\ast)}$  & -  \\
34F & GRP &  0.262 & A: & 1.25$\pm$0.29\\
70A & AGN I &  1.008 & A& -\\
82A & AGN I & 0.960  & $^{\ast)}$ & -  \\
104A& AGN II&  0.137 & ABCD & 1.89$\pm$0.49\\
128D& GRP &  0.031 & no & 1.08$\pm$0.36\\
\noalign{\smallskip}
131Z& CLUS &  0.205 & ABCD& 2.27$\pm$0.54\\
232A& star&   -  & ABCD, H$\beta$&1.55$\pm$0.88 \\
426A& AGN I&  0.788 & A:& 1.26$\pm$0.59\\
513~(34O) &AGN I & 0.761 & A& 1.39$\pm$0.72 \\
802A& star&    -    & BCD& -\\
\noalign{\smallskip}
815C& CLUS &  0.700 & AB & 2.18$\pm$0.53 \\
827B& AGN II&  0.249 & no & -\\
840D& GRP &  0.074 & ABCD, H$\beta$, H$\alpha$ & -\\
870~(36F) & AGN I  & 0.807 & A & 1.27$\pm$0.25\\
901A& AGN II&  0.205 & ABCD, H$\beta$& 1.84$\pm$0.40\\
      \noalign{\smallskip}
      \hline
      \end{tabular}
      \end{flushleft}
A) Ca H+K $\lambda$3934/3968, B) CH G $\lambda$4304, C) Mg I $\lambda$5175, D) Na I $\lambda$5890.\\
$^{\ast)}$ Lines are not covered by the spectrum (see Fig. \ref{spec1}).
      \end{table}

The UDS AGNs cover a redshift range from 0.080 to 4.45 with a median redshift z $ = $ 1.2, which is only slightly larger than that of the RDS AGN sample (z $ = $ 1.1).  In Paper III we compared the FWHM and the EW distributions of the RDS AGNs with those from several X-ray selected samples (e.g., the RIXOS sample of Puchnarewicz et al. \cite{Puch97}, the CRSS sample of Boyle et al. \cite{Boy97}) and from optical/UV selected AGN/quasar samples (e.g., the AGN samples of Steidel \& Sargent \& \cite{Stei91}, of Brotherton et al. \cite{Bro94}, and of Green \cite{Gre96}.) 

We have found a good agreement for the distributions of both broad and narrow emission lines. Slightly smaller mean EW values of the narrow emission lines are probably due to significant continuum emission from the host galaxies of several RDS AGNs.

To compare the results obtained from the RDS sample we have derived the mean FWHM and the mean EW for the UDS AGN sample, which contains in total 70 objects. Table \ref{line_comp} shows the comparison of emission line properties of the UDS AGN sample and the RDS AGN sample, which is a subsample of the UDS at higher X-ray flux. The first column gives the name of the most prominent emission lines. Column 2 marks the line component. The columns 3 and 4 (5 and 6) give the mean FWHM (EW), its 1$\sigma$ error, and the number of lines found in the RDS AGNs. The same data are shown for the UDS AGNs in columns 7 to 10. The very broad line components (FWHM$>$8000 km s$^{-1}$) have not been considered here.
The mean EW and FWHM of the two samples agree very well with each other, confirming the results from the comparison of the RDS AGN sample with other AGN/quasar samples from Paper III.  

For completenes Table \ref{gal_abs} shows the galaxy absorption lines of those UDS objects, that do not belong to the RDS sample (see Table 1 in Paper III for similar data for the objects in the RDS sample). Objects with $z>1.1$ are not listed, because the mentioned galaxy absorption lines lie outside the covered wavelength range.

The columns 1 and 2 contain the name and the class of the objects (AGN I -- type I AGN, AGN II -- type II AGN, GRP/CLUS -- galaxy group/cluster of galaxies). The redshift of the objects and the typical galaxy absorption lines found in the optical spectra are given in the columns 3 and 4. The entry ''no'' means that the absorption lines are not detected. Column 4 shows the 4000~\AA~break index, as defined by Bruzual (\cite{Bru83}), and its 1$\sigma$ error.

Seven out of the 11 new identified AGNs with $z<1.1$ show typical galaxy absorption lines in their optical spectra. Four of those AGNs have D(4000) values clearly larger than 1.0, which is an indication for a large continuum emission from their host galaxies (see Paper III for a detailed discussion). This is consistent with the fact 10 of these 11 AGNs have an absolute magnitude fainter than -22.0, well in the range covered by normal galaxies. 

\section{Near-infrared photometry}

A deep broad-band K$^{\prime}$ (1.9244--2.292 $\mu$m) survey of the Lockman Hole region has been performed with the Omega-Prime camera (Bizenberger et al. 1998) on the Calar Alto 3.5-m telescope in 1997 and in 1998. Some of the data were kindly provided to us by M. McCaughrean. About half of the ultradeep HRI pointing area is already covered. We planned to observe the remaining area in spring 2000, but weather conditions did not allow us to finish the $K^{\prime}$ survey. 

The camera uses a 1024$\times$1024 pixel HgCdTe HAWAII array with an image scale of 0.396 arcsec~pixel$^{-1}$, and covers a field-of-view of 6.7$\times$6.7 arcmin. A large number of background limited images were taken at slightly dithered positions. The total accumulated integration time of the combined images ranges from 45 to 70 minutes. The image reduction involves the usual standard techniques. The SExtractor package (Bertin \& Arnouts \cite{Ber96}) was used to detect the sources and to measure their fluxes. Point sources are well detected at the 5$\sigma$ level at K$^{\prime}$=19.7 mag in a 45 min (net) exposure.

   \begin{table}[t]
      \caption{Optical and NIRC-photometry of 14Z, 84Z, 486A, and 607 (36Z).}
      \label{red_sources}
      \begin{flushleft}
    \begin{tabular}{lcccc}
      \hline\noalign{\smallskip}
Filter  &  14Z & 84Z & 486A & 607 (36Z) \\
      \noalign{\smallskip}
      \hspace{0.1cm}(1) & (2) & (3) & (4) & (5)   \\\noalign{\smallskip}
      \hline \noalign{\smallskip}
$V$  & 25.4$\pm$0.3 & $>$25.0 & 24.9$\pm$0.1 & 24.2$\pm$0.1 \\
$R$  & 25.0$\pm$0.3 & $>$24.5-25.5& 24.4$\pm$0.1 & 24.1$\pm$0.2 \\
$I$  & 23.8$\pm$0.2 & 23.6$_{-0.2}^{+0.4}$ & 22.7$\pm$0.1 & 22.4$\pm$0.1 \\
$z$  & $>22.9$      & 22.5$\pm$0.3 & 21.4$\pm$0.1 &  -           \\
$J$  & 21.5$\pm$0.1 & 21.7$\pm$0.2 & 20.8$\pm$0.1 & 20.5$\pm$0.1 \\ 
$H$  & 20.5$\pm$0.1 & 20.0$\pm$0.1 & 19.8$\pm$0.1 & 20.2$\pm$0.2 \\
$K$  & 19.3$\pm$0.1 & 19.4$\pm$0.1 & 18.7$\pm$0.1 & 19.5$\pm$0.1 \\
       \noalign{\smallskip} \hline \noalign{\smallskip}
      \end{tabular}
      \end{flushleft}
      \end{table}

Four of the spectroscopically unidentified sources in this paper, 14Z, 84Z, 486A, and 607-36Z, were 
observed on UT 1999 December 15 and 16 under good seeing and photometric 
conditions using the facility Near-Infrared Imaging Camera (NIRC, 
Matthews \& Soifer \cite{Ma94}) on the Keck~I telescope.  NIRC has an 
image scale of 0\farcs15 per pixel, imaging onto a $256^2$ InSb detector 
for a 38\farcs4 square field of view.  Objects 14Z, 84Z, and 486A were 
observed for 10 minutes in each of the $zJHK$ bands; 607-36Z was 
observed for 5 minutes in the $JHK$ filters (the $z$ band data were 
saturated because the data were obtained near dawn).  The telescope 
was dithered in a random pattern every one minute of integration.  
The images were sky-subtracted and flatfielded, then stacked using 
integer pixel offsets.  Calibration onto the Vega flux scale was done using 
the Persson et al. (\cite{Pers98}) infrared standard stars.  

\begin{figure}[t]
     \begin{minipage}{86mm}
\psfig{figure=r_rmk_new.eps,bbllx=30pt,bblly=1pt,bburx=467pt,bbury=440pt,width=85mm,clip=}
     \end{minipage}
     \caption[]{$R-K^{\prime}$ colour versus optical R magnitudes for all objects in the Lockman Hole. Same symbols as for the X-ray sources in Fig. \ref{x_rdis}.Plus signs mark the remaining optically unidentified sources, where we show the brightest optical counterparts in the 80\% confidence level circle. All X-ray sources not covered by the $K^{\prime}$ survey so far are plotted at $R-K^{\prime}=0$ or 0.2.
Small dots show field objects not detected in X-rays.} %
     \label{rmk}
\end{figure}

Fig. \ref{rmk} presents the $R-K^{\prime}$ colour versus the $R$ magnitude for almost half of the objects of the UDS field.  We have detected all X-ray sources at the K$^{\prime}$ images, which are covered by the deep K$^{\prime}$ band survey in the Lockman Hole to date. There is a trend for the $R-K^{\prime}$ colour of the X-ray counterparts to increase to fainter $R$ magnitudes. Four of the optically unidentified X-ray sources have only very faint optical counterparts ($R>24$). The $K^{\prime}$ images of these sources (see Fig. 3 in Paper III) show relatively bright counterparts resulting
in very red colours ($R-K^{\prime}>5.0$). 

We have already argued in Paper III that such red counterparts of X-ray sources suggest either obscured AGNs or a high-z cluster of galaxies. The spectroscopic identification of five counterparts with $R-K^{\prime}>4.5$ confirms this so far (see Table \ref{id_tab}).
However, we cannot exclude that the very red objects are high-z quasars at $z>4.0$. The most distant X-ray selected quasar (817A) found to date (Schneider et al. \cite{Schn98}) shows also a relatively red colour ($R-K^{\prime}\sim4$). 

\section{Photometric redshift determination}

We have estimated photometric redshifts for the four unidentified sources
14Z, 84Z, 486A, 607 with broad-band photometry in several filters,
described in Sect. 4. Due to the hardness of 14Z, 84Z and 486A (see Table \ref{xray_tab}) and their large $R-K^{\prime}$ colours we argue that they are probably obscured AGNs. The photometric redshift of these objects is based on the assumption that their spectral energy distribution (SED) in the optical/near-infrared is due to stellar processes. 
If emission from an obscured AGN is contributing significantly at some
wavelength, the following results should be taken with caution.

We used a standard photometric redshift technique (see e.g., Cimatti et al. \cite{Cim99} and Bolzonella et al. \cite{Bolz00}).
In our version of the software
the templates consist of a set of synthetic spectra (Bruzual \&
Charlot \cite{Bru93}), with different star formation histories
and spanning a wide range of ages
(from $10^5$ to $2\times10^{10}$ yrs);
the basic set of templates includes only solar metallicity
and Salpeter's IMF.
The effect of IGM attenuation (Madau \cite{Mad95}),
extremely important at high redshift, is included, 
along with the effect of internal dust attenuation, 
using a dust--screen model and the SMC extinction law with E(B-V) ranging
from $0$ to $0.5$. 
In total 
$768$ synthetic spectra have been used.

\begin{figure}[hb]
     \begin{minipage}{87mm}
        \psfig{figure=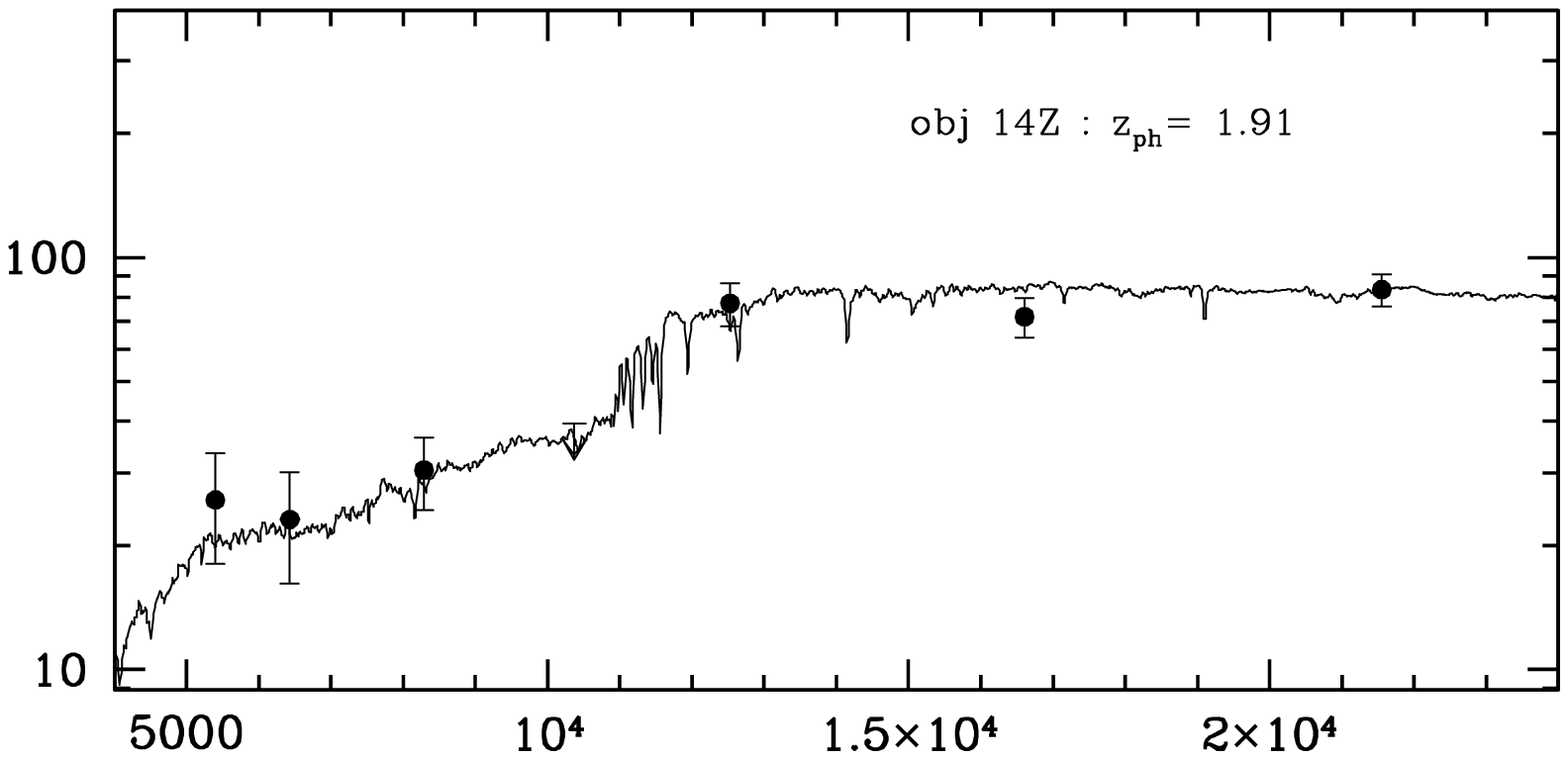,bbllx=73pt,bblly=433pt,bburx=531pt,bbury=655pt,width=87mm,clip=}
     \end{minipage}
     \begin{minipage}{87mm}
        \psfig{figure=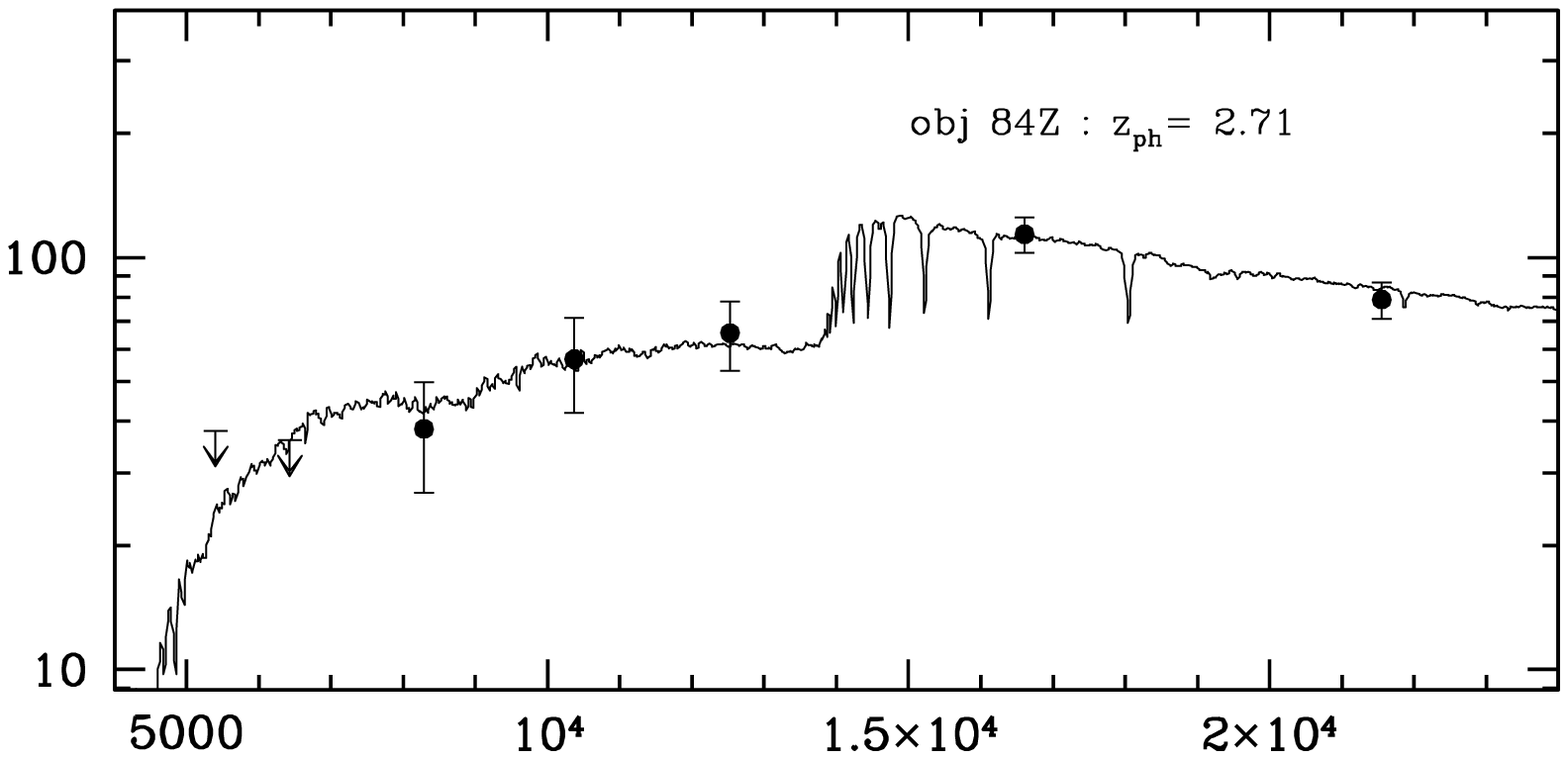,bbllx=73pt,bblly=433pt,bburx=531pt,bbury=655pt,width=87mm,clip=}
     \end{minipage}
     \begin{minipage}{87mm}
        \psfig{figure=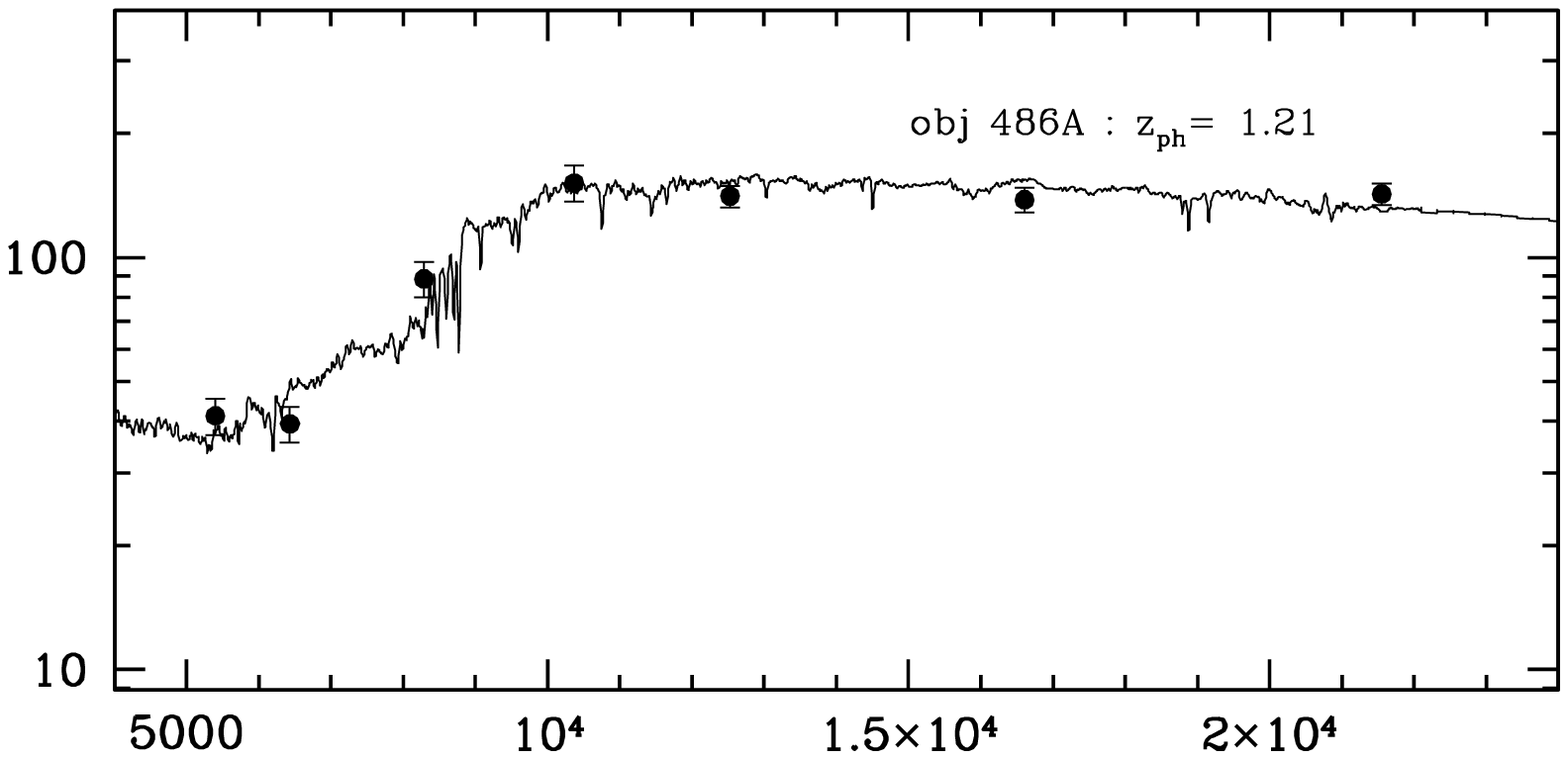,bbllx=73pt,bblly=433pt,bburx=531pt,bbury=655pt,width=87mm,clip=}
     \end{minipage}
     \begin{minipage}{87mm}
        \psfig{figure=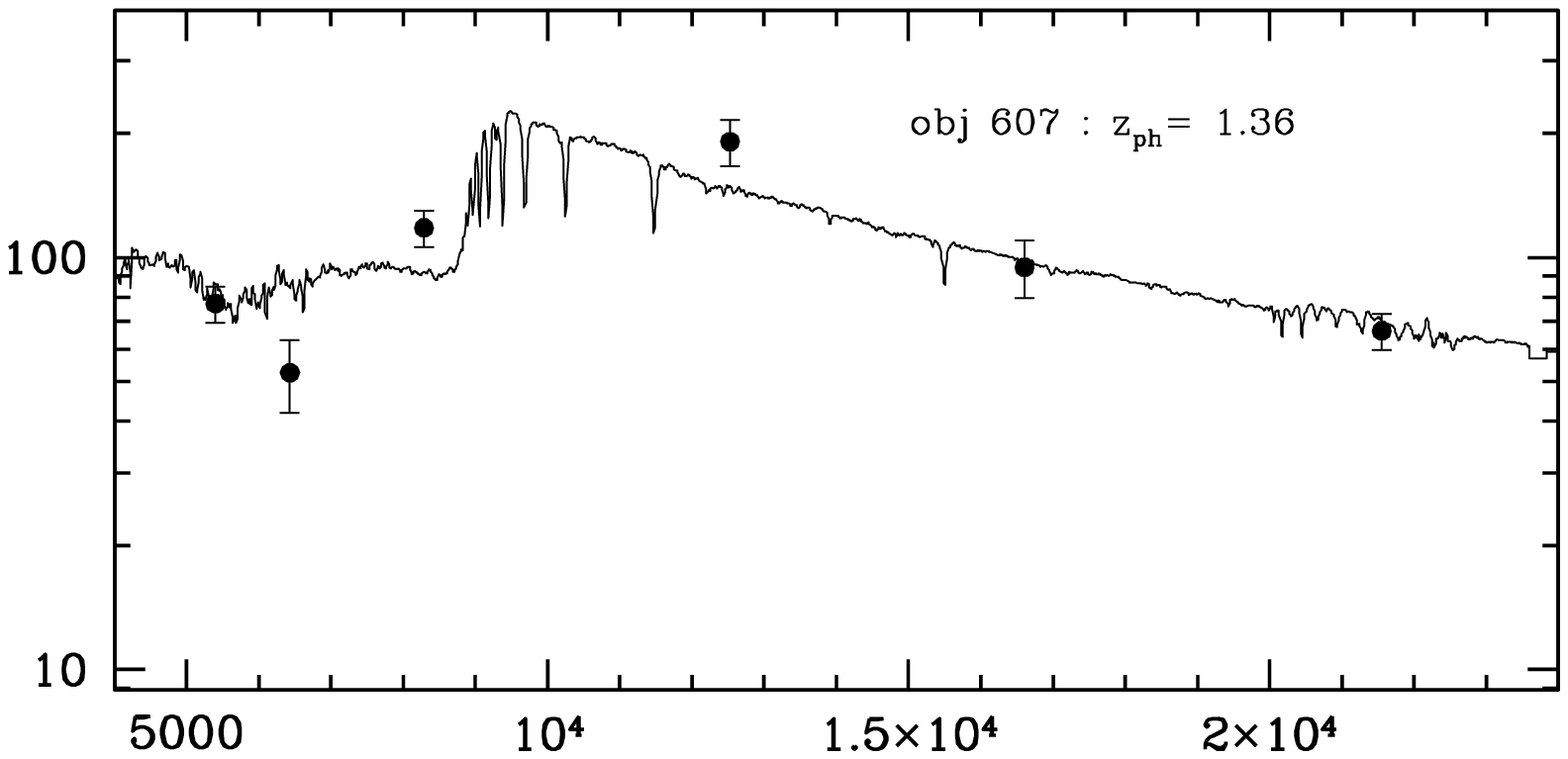,bbllx=73pt,bblly=433pt,bburx=531pt,bbury=655pt,width=87mm,clip=}
     \end{minipage}
     \caption[]{Galaxy template fits to the $VRIzJHK$ photometry of 14Z, 84Z, 486A, and 607 (36Z). The wavelength scale is in \AA ngstrom, the logarithm of the flux is in units of 10$^{-20}$ erg s$^{-1}$ cm$^{-2}$ \AA $^{-1}$.}
               \label{red_phot}
\end{figure}
%

The ``best" photometric redshift ($z_{\rm phot.}$) for each galaxy is
computed by applying a standard, error-weighted $\chi^2$ minimization
procedure.
Moreover we have computed error bars to $z_{\rm phot.}$ corresponding to
90\% confidence levels,
computed by means of the $\Delta \chi^2$ increment for
a single parameter (Avni \cite{Av76}).
The observed spectral energy distribution (SED) of each objects,
obtained from broad--band photometry in several filters
($V,I$ from 8K UH, $R$ from Keck+LRIS and $z,J,H,K$ from
Keck+NIRC), is compared to our set of template spectra.
The $V,I$ 8K UH observations are described by Wilson et al. (\cite{Wil96}).

In agreement with their very red colours ($4.6 < R-K < 5.7$) we obtained
relatively high $z_{\rm phot.}$, ranging from $1.21<z_{\rm phot.}<2.71$,
for all sources (Fig. \ref{red_phot}).
For objects 14Z and 486A we estimate
$z_{\rm phot.}=1.94^{+0.18}_{-0.10}$ and
$z_{\rm phot.}=1.21^{+0.10}_{-0.14}$, respectively. Both observed SEDs
are consistent with an old stellar population (age=$2.5 \div 5$ Gyrs),
while the content of dust is badly constrained because of the absence of
$U$ and $B$ band photometry.

The formal best estimate for redshift of object 607 (36Z) is 
$z_{\rm phot.}=1.36^{+0.07}_{-0.12}$, but the resulting fit is very poor;
the upturn in the $V$ photometry at $\lambda<6000$
indicates the presence of a young stellar population
(0.3 Gyrs) and the absence or a low dust extinction or the presence of an underlying AGN component. The photometric redshift of 607 (36Z) is therefore very uncertain. 
Data in bluer bands ($U$ and $B$) and combined AGN and galaxy templates would
probably be needed for a more reliable redshift estimate for this object.
We therefore have not included the value of 607 in Table \ref{id_tab}.

Object 84Z shows a quite clear break between $J$ and $H$ photometry,
consistent with a best fit model at
$z_{\rm phot.}=2.71^{+0.29}_{-0.41}$, while the decreasing flux towards shorter wavelength indicates the presence of a moderate dust content ($E(B-V)=0.3$) in a young stellar population (age=$0.1$ Gyrs). 
As seen in Fig \ref{red_phot}, the resulting fit is very good, with a $\chi^2$
value of the order of unity. However, due to the relatively large magnitude
errors, especially in the $I$ and $z$ bands, also lower redshifts (down to
z $\sim$ 1.5) would be statistically acceptable.

\section{Discussion and conclusion}

We have presented the nearly complete optical identification of 94 X-ray sources
with 0.5-2.0 keV X-ray fluxes larger than $1.2 \cdot 10^{-15}$ erg~s$^{-2}$ from the Ultra Deep ROSAT Survey in the Lockman region. Highly accurate HRI positions, deep Keck $R$ and Palomar $V$ images, and high signal-to-noise ratio Keck spectra allow a reliable identification of 85 (90\%) X-ray sources. 

Table \ref{ID_sum} shows the spectroscopical identification summary of the UDS, compared with that of the RDS. As seen in the table, the population of optical counterparts in the UDS has not changed from that of the RDS, with the large majority of the identifications (75\% of the X-ray sources)
being identified with AGN.
The ratio between AGN I (57 objects) and AGN II
(13 objects) is greater than 4, although it could decrease to about 3 if some, or most, of the 8 remaining spectroscopically unidentified X-ray sources
are type II AGN (see Sect. 6).
The second most abundant class of objects is constituted by
groups and/or clusters of galaxies (10), followed by galactic stars (5).

We have spectroscopically identified only one source in the entire sample with a ''normal'' emission line galaxy (53A). 
The ASCA Deep Survey of the Lockman Hole by Ishisaky et al. (\cite{Ish01}) suggests an obscured AGN also in this case.

   \begin{table}[t]
     \caption[]{Spectroscopical identification summary of the UDS and of the RDS.}
      \label{ID_sum}
      \begin{flushleft}
      \begin{tabular}{lrcrc}
      \hline\noalign{\smallskip}
             & \multicolumn{2}{c}{UDS} & \multicolumn{2}{c}{RDS}\\
Object class & N\hspace*{0.15cm} & Content & N\hspace*{0.15cm} & Content \\\noalign{\smallskip} 
\hspace*{0.6cm}(1) & (2) & (3) & (4) & (5)\\ \noalign{\smallskip}
      \hline\noalign{\smallskip}
      \noalign{\smallskip}
AGN type I           &  57  & 61\% &31 &62\%\\
AGN type II$^{\ast)}$ &  13  & 14\%&10 &20\%\\
Cluster/groups       &  10  & 11\% &3&6\%\\
Stars                &   5  &  5\% &3&6\%\\
Galaxies             &   1  &  1\% &1&2\%\\
Unidentified sources$^{\ast\ast)}$ &   8  &  8\% &2&4\%\\ \noalign{\smallskip}
      \hline\noalign{\smallskip}
Total                &  94  & -& 50 & -\\
     \noalign{\smallskip}
      \hline
      \end{tabular}
      \end{flushleft}
$^{\rm \ast)}$ including the objects 28B and 59A with broad Balmer emission lines but large Balmer decrements.\\
$^{\rm \ast\ast)}$ including the very red AGN type II objects 14Z, 84Z and 486A, which are still
spectroscopically unidentified.
   \end{table}

These results clearly show that the fraction of AGN as optical/infrared
counterparts of faint X-ray sources in the 0.5-2.0 keV energy band remains
high down to a limiting flux of 1.2$ \cdot 10^{-15}$ erg~cm$^{-2}$s $^{-1}$,
a factor $\sim$4.6 times fainter than the RDS survey, and confirm that the
soft X-ray background is dominated by the emission of type I AGN, and that the increased sensitivity of the UDS has not revealed an increase in the fraction 
of type II AGN.

In the following we briefly summarize, in turn, some of the properties of the
objects in the three main classes of optical counterparts.\\

a) Type I AGN\\

$\bullet $ The emission line properties of the 57 type I AGN in the UDS survey
are consistent with those of other brighter X-ray selected samples
and optical/UV selected samples.

$\bullet $ The X-ray luminosity of these objects covers the range
43 $ < $ log $ L_{\rm x} < $ 45, confirming  
that most of the contribution to the X-ray background from AGN I is due to moderately powerful objects.

$\bullet $ The X-ray and optical luminosity of the type
I AGN are reasonably well correlated, with an average value corresponding
to $f_{\rm x}/f_{\rm v} \sim~$ 1 (see Fig. \ref{logLxMv}).

$\bullet$ The UDS contains the most distant X-ray selected quasar ($z=4.45$) found to date (Schneider et al. \cite{Schn98}). 

$\bullet $ The surface density of type I AGN in the HRI defined sample,
which is the deepest part of the UDS survey (40 objects in 0.126 sq.deg.,
corresponding to 317 $\pm$ 50 objects per sq.deg.) is higher than any
reported surface density based on spectroscopic samples for this class of
objects. This confirms the very high efficiency of X-ray
selection in detecting this kind of objects (see discussion in Zamorani et
al. \cite{Zam99} for a comparison of the relative
efficiencies of X-ray and optical selections of type I AGN).\\

\begin{figure}[th]
     \begin{minipage}{87mm}
       \psfig{figure=logLx_Mv_new2.eps,bbllx=121pt,bblly=141pt,bburx=481pt,bbury=651pt,width=87mm,angle=-90,clip=}
     \end{minipage}
     \caption[]{The logarithm of the 0.5-2.0 keV X-ray luminosity is plotted versus the absolute magnitude $M_{\rm v}$ of the optical counterparts marked with different symbols; filled circles -- type I AGNs, open circles -- type II AGNs, open squares -- groups/clusters of galaxies, hexagon -- galaxy, and crosses -- very red sources with known photometric redshift (type II AGNs). The solid line corresponds to the X-ray to optical flux ratio $f_{\rm x}/f_{\rm v}=1$ typical for AGNs. All except four AGNs have X-ray luminosities larger than 10$^{43}$ erg~s$^{-1}$ (above the dotted line).  }
               \label{logLxMv}
\end{figure}
%

b) Type II AGN and unidentified X-ray sources\\

Following Schmidt et al. (1998; Paper II) and Lehmann et al. (2000; Paper
III), we have adopted a variety of indicators (e.g. optical diagnostic
diagrams, presence of [Ne~V] and/or strong [Ne~III] forbidden lines in the spectrum,
large X-ray luminosity) to classify an object as AGN II. In addition, we have
put in this class also two ``intermediate'' objects (28B and 59A), which,
although having broad Balmer lines, show a clear indication of significant
absorption as suggested by their large Balmer decrement.

\begin{figure}[htb]
     \begin{minipage}{86mm}
\psfig{figure=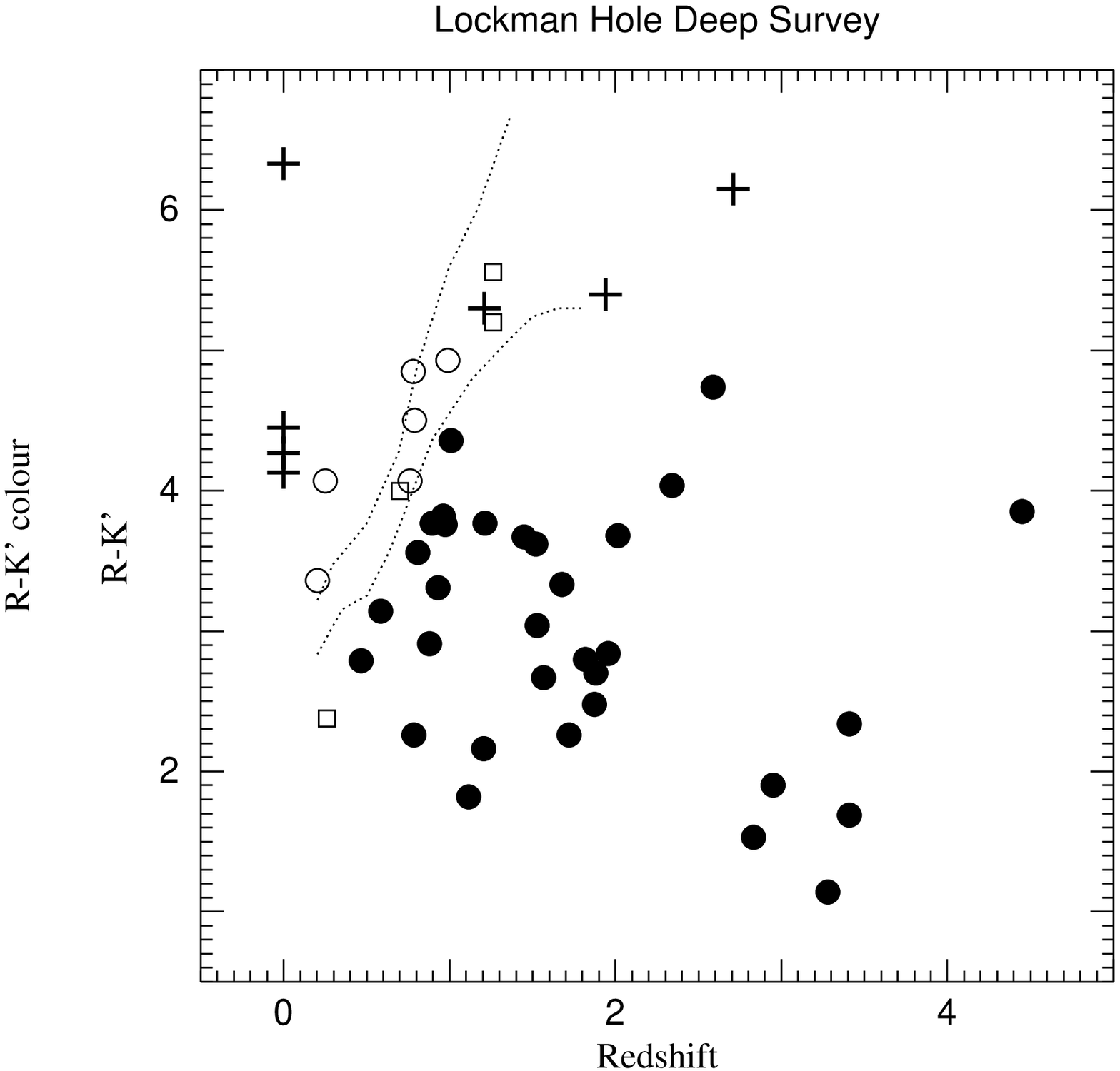,bbllx=30pt,bblly=1pt,bburx=485pt,bbury=440pt,width=85mm,clip=}
     \end{minipage}
     \caption[]{$R-K^{\prime}$ colour versus redshift for those X-ray sources in the Lockman Hole with available $K^{\prime}$ band photometry. Same symbols are shown as in Fig. \ref{logLxMv}. For the objects 14Z, 84Z, and 486A we have used photometric redshifts (see Sect. 6). The dotted lines are from Steidel and Dickinson (\cite{Stei94}) corresponding to unevolved spectral models for E (upper) and S$_{\mathrm{b}}$ (lower) galaxies from Bruzual and Charlot (\cite{Bru93}).}
     \label{zrmk}
\end{figure}

$\bullet$ While the colours of type I AGN are relatively blue,
type II AGN and groups/clusters of galaxies show on average much redder
colours. The location of type II AGN in the redshift - $(R-K^{\prime})$ diagram  (see Fig. \ref{zrmk}) occupies the same region as expected for elliptical and spiral galaxies.
The colours of type II AGN appear to be dominated by the light from their
host galaxies. The optical spectra of type II AGN confirm that a
substantial fraction of their emission originates in their host galaxies
(see Sect. 4.3). Two type II AGN (12A and 117Q), with spectroscopic redshifts
0.990 and 0.780, have very red colours ($R-K^{\prime}\sim5.0$). Both sources
are detected with the PSPC and show hard spectra. It is therefore likely that these objects
are significantly absorbed in X-ray.

$\bullet $ Most of the type II AGN (9 out of 13) have X-ray luminosities
in the relatively narrow range 43 $ < $ log $ L_{\rm x} < $ 43.7 and are approximately
consistent, although toward low luminosities, with the $L_{\rm x} - M_{\rm v}$
relation defined by the type I AGN  (see Fig. \ref{logLxMv}). The X-ray luminosity
distributions of the two classes of AGN are significantly different:
we have 37 type I and no type II AGN with log $L_{\rm x} >$ 43.7, but 
19 type I and 13 type II AGN with log $L_{\rm x} <$ 43.7.

$\bullet $ Four objects (59A, 104A, 827B, 901A) have X-ray luminosities
significantly smaller than all the other AGN (log $L_{\rm x} \sim$ 41.5), with
correspondingly low $f_{\rm x}/f_{\rm v}$ ratio. All of them are at low redshift ($z<0.25$).

$\bullet $ The different distributions in X-ray luminosity of type II and
type I AGN translates into significantly different distributions in redshift.
For example, up to $z=0.25$ we have 5 type II and no type I AGN.
The ratio between the type I and type II AGN, which is $\sim$ 4.3
for the entire sample, is $\sim$ 0.3 up to $z=0.5$, $\sim$ 0.4 up to
$z=0.75$ and $\sim$ 1 up to $z=1$.

$\bullet $ These trends of the ratio
between the type I and type II AGN (increasing with redshift and/or
luminosity) would remain qualitatively similar even if most of the
8 spectroscopically unidentified sources will turn out to be, as we have argued in Sect. 6,
high redshift type~II AGN. For three of these sources, assuming that their
optical/near-infrared SED is mainly due to stellar processes, we have
determined photometric redshifts ranging from 1.22 to 2.71. These redshifts
lead to X-ray luminosities in the
0.5-2.0 keV band of up to 10$^{44} $ erg~s$^{-1}$, which is in the regime of
typical QSO X-ray luminosities. 

Even in this case, however, the number of high luminosity type II AGN in our sample appears to be smaller than that expected in the simplest version of the unified models which has been used so far for the AGN synthesis models of the X-ray background (see e.g., Comastri et al. \cite{Co95} and Gilli et al.  \cite{Gi99}). 

This is consistent with the fact that until now only a few hard X-ray
selected type II AGN at high luminosity are known (Akiyama et al. \cite{Aki00},
Nakanishi et al. \cite{Na00}). In addition, the recent 1-7 keV ASCA Deep
Survey in the Lockman Hole suggests a deficit of high luminous absorbed
sources at $z\sim 1-2$. Unfortunately, all these samples, including
ours, are based on a very limited number of sources, so that a strong
conclusion is not warranted yet. For example, Gilli et al. (\cite{Gi01}) have
recently shown that the redshift distribution of type I and type II
AGN for a sub-sample of the UDS sources is statistically consistent
with models in which the fraction of obscured AGN is constant with
luminosity. Only significantly larger spectroscopic samples of hard
X-ray selected AGN from Chandra and XMM-Newton observations can
fully clarify this issue. For example, 
the PV-observation of the Lockman Hole with XMM-Newton has recently
revealed a large fraction of very red sources in the 2-10 keV energy band,
which are probably heavily obscured AGNs (Hasinger et al.
\cite{Has01}, see also Barger et al. \cite{Ba01} for similar finding in the Chandra observation of the Hawaii Deep Survey Field SSA13), but no spectroscopic identifications for these objects currently exist.\\




c) Groups and clusters of galaxies\\

$\bullet$ Nine out of the ten X-ray sources identified with groups
and/or clusters of galaxies are extended either in the HRI or in the
PSPC images or both. The point-like source 815 is classified as a cluster on the
basis of the optical data: three galaxies within a few arcseconds of the X-ray position have the same redshift ($z=0.700$).

$\bullet$ Nine out of ten objects have X-ray luminosities in the range
41.5 $ < $ log $L_{\mathrm x} < $ 43.5 and cover the redshift range 0.20 - 1.26. The median redshift of this sample of groups is $\sim$ 0.5.

$\bullet$ The faintest X-ray group in our sample (identified with the
X-ray source 840) has a very low X-ray luminosity (log $L_{\mathrm x}$ = 40.7).
In this case, given the small redshift ($z=0.074$) we cannot exclude the 
possibility 
that the X-ray emission is due to a single galaxy or to the sum of the
emission of a few galaxies in the group. 

$\bullet$  We have detected one of the most distant X-ray selected cluster
of galaxies at $z=1.263$ known to date (Hasinger et al. \cite{Has98}; Paper IV, Thompson et al. \cite{Tho01}).


\begin{acknowledgements}
The ROSAT project is supported by the Bundesministerium f\"ur
Bildung, Forschung und Technologie (BMBF/DFR) and the Max-Planck-Gesellschaft.

Data presented herein were obtained at the W.M. Keck Observatory, which is operated as a scientific partnership among the California Institute of Technology,
the University of California and the National Aeronautics and Space Administration.  The Observatory was made possible by the generous
financial support of the W.M. Keck Foundation. We thank J. Oke and J. Cohen for their efforts in the contruction of LRIS, and B. Schaefer for organizing service observing.

I.L. was visiting astronomer at the German-Spanish Astronomical Centre, Calar Alto, operated by the Max-Planck-Instiute for Astronomy, Heidelberg, jointly with the Spanish National Commission for Astronomy. I.L. thanks for the support by the DFG travel grant (HA 1850/11-1). 

Partial support for this work was provided by National Science Foundation
grant AST99-00703 and by NASA through grant number GO-08194.01-97A from
the Space Telescope Science Institute, which is operated by the Association
of Universities for Research in Astronomy, Inc., under NASA contract
NAS5-26555 (DPS).

G.Z. acknowledges partial support by the Italian Space Agency (ASI) under ASI
contract ARS-98-119 and by the Italian Ministry for University and Research
(MURST) under grant Cofin98-02-32.

Part of this work was supported by the German {\it Deutsches Zentrum f\"ur Luft- und Raumfahrt, DLR} project numbers 50 OX 9801 and 50 OR 9908.

We thank our referee R.S. Warwick for his helpful comments and suggestions.
\end{acknowledgements}

\clearpage

\appendix
\section{Optical images and spectra}

The optical images of all X-ray sources previously not contained in the RDS sample (see Lehmann et al. \cite{Leh00a}), are presented in Fig. \ref{spec1}. The $R$ images show a 40$''$ $\times$ 40$''$ field of view, in most cases centered on the position of the X-ray source. The field size of the $V$ images is $\sim$60$''$ $\times$ 60$''$. Overplotted are 80\% confidence level error circles of the PSPC detection (large) and of the HRI detection (small). VLA 20 cm detections from DeRuiter et al. (\cite{deRui97}) are marked with 2$\sigma$ error boxes.

In addition, Fig. \ref{spec1} shows the low-resolution Keck spectra of the optical counterparts of the X-ray sources.  The location
of UV emission lines (e.g., Ly$\alpha$~$\lambda$1216, Si~IV~$\lambda$1397, C IV~$\lambda$1548 , C III]~$\lambda$1908, Mg II~$\lambda$2798) and the lines of the Balmer series as well as narrow forbidden lines of Neon and Oxygen (e.g., [Ne~V]~$\lambda$3426, [O~II]~$\lambda$3727) are indicated.   Furthermore
there are marked typical galaxy absorption lines (Ca~H+K~$\lambda$$\lambda$3934/3968, CH~G~$\lambda$4304, Mg~I~$\lambda$5175, Na~I~$\lambda$5890).



\begin{figure*}
     \begin{minipage}{177mm}
\psfig{figure=fxig1_uds_new.ps,bbllx=66pt,bblly=55pt,bburx=556pt,bbury=664pt,width=177mm}
     \end{minipage}
     \caption[]{$R$-band/$V$-band images and Keck LRIS spectra of the new optical counterparts. North is up, and east to the left. The small and large circles show the ROSAT HRI and PSPC error circles (80 \% error radius). The 2$\sigma$ error box indicates a VLA 20 cm detection. The wavelength is given in \AA ngstrom, the flux scale is normalized such that one count corresponds to an AB magnitude 28 (f$_{\mathrm \nu}=2.29 \cdot 10^{-31}$ erg s$^{-1}$ cm$^{-2}$ Hz$^{-1}$).
     Regions with bad atmospheric correction are 
     marked with ''x''.  Residuals of night sky emission lines are indicated with ''s''.
 }
     \label{spec1}
\end{figure*}
\addtocounter{figure}{-1}

%
\begin{figure*}
     \begin{minipage}{177mm}
\psfig{figure=fixg2_uds_new.ps,bbllx=66pt,bblly=55pt,bburx=556pt,bbury=664pt,width=177mm}
     \end{minipage}
     \caption[]{(continued) 
 The spectrum of source 232 was taken with the Boller \& Chivens Cass Twin spectrograph at the Calar Alto 3.5m telescope. The asterisk symbol in the image of source 128 marks the most probable counterpart (see Sect. 4.1.).
 }
     \label{spec2}
\end{figure*}
\addtocounter{figure}{-1}
%
\begin{figure*}
     \begin{minipage}{177mm}
\psfig{figure=fixg3_uds_new2.ps,bbllx=66pt,bblly=55pt,bburx=556pt,bbury=664pt,width=177mm}
     \end{minipage}
     \caption[]{(continued)}
    \label{spec3}
\end{figure*}
\addtocounter{figure}{-1}

%
\begin{figure*}
     \begin{minipage}{177mm}
\psfig{figure=fixg4_uds_new2.ps,bbllx=66pt,bblly=55pt,bburx=556pt,bbury=664pt,width=177mm}
     \end{minipage}
    \caption[]{(continued)}
    \label{spec4}
\end{figure*}
%
\addtocounter{figure}{-1}

%
\begin{figure*}
     \begin{minipage}{177mm}
\psfig{figure=fixg5_uds_new.ps,bbllx=66pt,bblly=296pt,bburx=556pt,bbury=400pt,width=177mm}
     \end{minipage}
    \caption[]{(continued)}
    \label{spec4}
\end{figure*}
%
\hfill
%
%
   \begin{table*}[b]
      \caption{FWHM of UV emission lines in [km s$^{-1}$], corrected for instrumental resolution.}
      \label{FWHM_UV}
      \begin{flushleft}
      \begin{tabular}{lc@{\hspace*{3mm}}c@{\hspace*{3mm}}rrrrrr}
      \hline\noalign{\smallskip}
name  &$z$&Ly${\alpha}$~$\lambda$1216 & Si IV~$\lambda$1397 &C IV~$\lambda$1548 & C III]~$\lambda$1908 & Mg II~$\lambda$2798 & [Ne V]~$\lambda $3426 \\
\noalign{\smallskip}
(1) & (2) & (3)\hspace{0.5cm} & (4)\hspace{0.6cm} & (5)\hspace{0.5cm} & (6)\hspace{0.6cm} & (7)\hspace{0.6cm}& (8)\hspace{0.6cm} \\ \noalign{\smallskip}
\hline\noalign{\smallskip}
5A   &1.881 & \point & \point  & 4450$\pm$70  & 2160$\pm$60  & \point  & \point \\
13A  &1.879 & \point & \point & \point & 11970$\pm$170$^{v}$  & 14470$\pm$220$^{\ast\ast}$  & \point \\
15A  &1.447 & \point & \point & \point & 4700$\pm$150 &  2150$\pm$50  & \point \\
18Z  &0.993 & \point & \point & \point & \point & 3670$\pm$130  & 310$\pm$180  \\
33A  &0.974 & \point & \point & \point & \point & 5420$\pm$130  & 1060$\pm$90  \\ 
\noalign{\smallskip}
39B  &3.269 & $<$810  & 7490$\pm$250  & 1780$\pm$40  & \point & \point & \point \\
     &      & 3370$\pm$70$^{b}$ &        & 7390$\pm$90       &        &        &        \\
70A$^{\ast}$  &1.008 & \point & \point & \point & \point & 7820$\pm$450  & \point \\ 
75A  &3.406 & 430$\pm$10  & 4560$\pm$70  & 1580$\pm$250  & \point & \point & \point \\
     &      & 1860$\pm$30$^{b}$ &  & 5240$\pm$70  &        &        &        \\
\noalign{\smallskip}
80A  &3.407 & $<$650  & 3600$\pm$260$^{m}$ & 2250$\pm$30  & \point & \point & \point \\
     &      & 2980$\pm$50$^{b}$ &        &        &        &        &        \\
82A  &0.960 & \point & \point & \point & \point & 3830$\pm$220  & \point \\ 
120A &1.573 & \point & \point & 5060$\pm$50  & 6130$\pm$80  & 4690$\pm$50  & \point \\
151B &1.201 & \point & \point & \point & \point & 5690$\pm$130  & \point \\
\noalign{\smallskip}
426A &0.788 & \point & \point & \point & \point & 1480$\pm$220  & 1000$\pm$200$^{m}$  \\
428E &1.539 & \point & \point & 4890$\pm$80  & 4140$\pm$220  & 5760$\pm$280$^{m}$ & \point \\
477A$^{\ast}$ &2.943 & 3440$\pm$30  & 4440$\pm$50  & 4690$\pm$20  & 5250$\pm$40  & \point & \point \\ 
513~(34O)  &0.761 & \point & \point & \point & \point & 2010$\pm$130  & \point \\
634A &1.546 & \point & \point & \point & \point & 3230$\pm$320  & \point \\
\noalign{\smallskip}
801A &1.675 & \point & \point & \point & 8480$\pm$340$^{v}$  & 4100$\pm$90  & \point \\
804A &1.214 & \point & \point & \point & \point & 2720$\pm$50  & 320$\pm$30  \\
805A &2.586 & 8310$\pm$210$^{v}$ & 4460$\pm$280  & 1980$\pm$110$^{b}$        & 1420$\pm$100 &  \point     &  \point     \\
     &      &        &                           & 11040$\pm$450$^{v}$ &            &              &             \\
817A &4.450 & $<$530  & 450$\pm$60 & \point & \point & \point & \point \\ 
\noalign{\smallskip}
     &      & 1720$\pm$80$^{b}$  &          &        &        &        \\
821A &2.310 & 2390$\pm$60$^{b}$  & 3360$\pm$240  & 2190$\pm$90$^{b}$  & \point & \point & \point \\
     &      & 8390$\pm$330$^{v}$ &               & 10120$\pm$300$^{v}$ & \point & \point & \point \\
828A &1.285 & \point & \point & \point & 4050$\pm$290  & 7640$\pm$150  & \point  \\
832A &2.735 & \point & 9560$\pm$110$^{v}$  & 570$\pm$50  & 2510$\pm$120  & \point & \point \\
\noalign{\smallskip}
     &      &        &        & 4020$\pm$80  &        &        &        \\
837A &2.018 & \point & \point & 4140$\pm$140  & 1090$\pm$80  & 920$\pm$40 & \point \\
861A &1.849 & \point & \point & 5870$\pm$130  & 7440$\pm$150  & \point  & \point \\ 
870~(36F)  &0.807 & \point & \point & \point & \point & 3080$\pm$180  & \point \\
901A &0.205 & \point & \point & \point & \point & \point & $<$920  \\
       \noalign{\smallskip}
      \hline
      \end{tabular}
      \end{flushleft}
$^{\rm n)}$ Narrow, $^{\rm b)}$ broad and $^{\rm v)}$ very broad component (FWHM$>$8000 km s$^{-1}$) of the emission line.\\
$^{\rm m)}$ Marginal line detection ($2-3\sigma$),  
$^{\rm \ast)}$ More than one spectrum available,
$^{\rm \ast\ast)}$ Poor Gaussian fit ($\chi_{\mathrm red} \sim 20$).\\
      \end{table*}
%

\section{FWHM and EW tables}

Tables \ref{FWHM_UV}--\ref{EW_opt} contain the instrumental corrected FWHM in km~s$^{-1}$ and the rest frame EW in \AA~ of the UV/optical emission lines from the newly identified optical counterparts of the X-ray sources.
The first two columns give the name of the objects and their mean redshifts. The FWHM values in columns 3 to 8 and 3 to 9 of the Tables  \ref{FWHM_UV} and  \ref{FWHM_opt} are rounded to the nearest integer that can be divided 10. The EW values given in columns 3 to 8 and 3 to 9 of the Tables  \ref{EW_UV} and  \ref{EW_opt} have been rounded to the tenths. The errors of the FWHM and EW values are propagated 1$\sigma$ errors provided by the fitting algorithm.

\newpage 

%
%
   \begin{table*}[t]
      \caption{Rest frame EW of UV emission lines in [\AA].}
      \label{EW_UV}
      \begin{flushleft}
      \begin{tabular}{l@{\hspace*{4mm}}c@{\hspace*{4mm}}rrrrrr}
      \hline\noalign{\smallskip}
name &$z$&Ly${\alpha}$~$\lambda$1216 & Si IV~$\lambda$1397 &C IV~$\lambda$1548 & C III]~$\lambda$1908 & Mg II~$\lambda$2798 & [Ne V]~$\lambda $3426 \\
\noalign{\smallskip}
(1) & (2) & (3)\hspace{0.5cm} & (4)\hspace{0.6cm} & (5)\hspace{0.5cm} & (6)\hspace{0.6cm} & (7)\hspace{0.6cm}&(8)\hspace{0.6cm} \\ \noalign{\smallskip}
\hline\noalign{\smallskip}
5A   &1.881  & \point & \point & 25.9$\pm$1.1  & 9.0$\pm$0.6  & \point  & \point \\
13A  &1.879 & \point & \point & \point & 17.7$\pm$0.6$^{v}$  & 45.7$\pm$2.1$^{\ast\ast}$  & \point \\
15A  &1.447 & \point & \point & \point & 16.3$\pm$1.3  & 17.7$\pm$1.0  & \point \\
18Z  &0.993 & \point & \point & \point & \point & 18.5$\pm$1.6  & 4.2$\pm$1.1  \\
33A  &0.974 & \point & \point & \point & \point & 14.6$\pm$0.8  & 3.3$\pm$0.5  \\ 
\noalign{\smallskip}
39B  &3.269 & 10.3$\pm$0.5  &  11.7$\pm$0.9  & 9.2$\pm$0.5  & \point & \point & \point \\
     &    & 26.9$\pm$1.8$^{b}$  &                & 39.2$\pm$1.7 & & & \\
70A  &1.008 & \point & \point & \point & \point & 47.2$\pm$7.9  & \point \\ 
75A  &3.406 & 2.7$\pm$0.1  & 4.3$\pm$0.2  & 3.2$\pm$0.1  & \point & \point & \point \\
     &    & 3.5$\pm$0.1$^{b}$  &        & 8.7$\pm$0.4  & \point & \point & \point \\
\noalign{\smallskip}
80A  &3.407 & 11.0$\pm$1.3  & 49.0$\pm$17.2$^{m}$  & 72.5$\pm$5.3  & \point & \point & \point \\
     &    & 65.3$\pm$5.8$^{b}$  &  &  & & & \\
82A  &0.960 & \point & \point & \point & \point & 40.8$\pm$10.0  & \point \\ 
120A &1.573 & \point & \point & 29.7$\pm$0.9  & 22.7$\pm$0.8  & 33.6$\pm$1.0  & \point \\
151B &1.201 & \point & \point & \point & \point & 44.5$\pm$2.8  & \point \\
\noalign{\smallskip}
426A &0.788 & \point & \point & \point & \point & 6.7$\pm$1.9  & 7.1$\pm$2.5$^{m}$  \\
428E &1.539 & \point & \point & 48.0$\pm$2.9  & 25.7$\pm$3.9  & 665.4$\pm$368.6$^{m}$  & \point \\
477A &2.943 & 42.8$\pm$0.9  & 12.9$\pm$0.4  & 26.1$\pm$0.4  & 28.4$\pm$0.5  & \point & \point \\
513~(34O)  &0.761  & \point & \point & \point & \point & 37.9$\pm$6.7  & \point \\
634A &1.546 & \point & \point & \point & \point & 35.1$\pm$10.3  & \point \\
\noalign{\smallskip}
801A &1.675 & \point & \point & \point & 29.2$\pm$3.4$^{v}$  & 20.7$\pm$1.2  & \point \\
804A &1.214 & \point & \point & \point & \point & 28.2$\pm$1.5  & 9.1$\pm$0.6  \\
805A &2.586 & 103.3$\pm$7.5$^{v}$  & 21.5$\pm$4.1 & 15.9$\pm$2.6$^{b}$   & 18.0$\pm$3.3  &  \point     &  \point     \\
     &      &             &                  & 59.6$\pm$9.2$^{v}$ &        &              &             \\
817A &4.450 & 8.6$\pm$1.1  & 10.5$\pm$1.7 & \point & \point & \point & \point \\ 
\noalign{\smallskip}
     &    & 16.1$\pm$1.6$^{b}$ &              &        &        &        & \\
821A &2.310 & 30.7$\pm$2.4$^{b}$  & 9.6$\pm$1.7  & 19.7$\pm$2.9$^{b}$  & \point & \point & \point \\
     &    & 36.9$\pm$3.2$^{v}$  &              & 84.8$\pm$10.8$^{v}$ &        & & \\
828A &1.285 & \point & \point & \point & 4.0$\pm$0.7  & 19.3$\pm$0.9  & \point  \\
832A &2.735 & \point & 101.9$\pm$4.9$^{v}$  & 7.4$\pm$0.6  & 10.9$\pm$1.0 & \point & \point \\
\noalign{\smallskip}
     &      &  &   & 27.3$\pm$2.1  &  &  & \\
837A &2.018 & \point & \point & 47.2$\pm$5.5  & 8.2$\pm$1.4  & 54.4$\pm$10.3 & \point \\
861A &1.849 & \point & \point & 28.5$\pm$2.0  & 32.2$\pm$1.8  & \point  & \point \\
870~(36F) &0.807 & \point & \point & \point & \point & 7.1$\pm$1.0  & \point \\
901A &0.205 & \point & \point & \point & \point & \point & 4.6$\pm$0.4  \\
       \noalign{\smallskip}
      \hline
      \end{tabular}
      \end{flushleft}
$^{\rm n)}$ Narrow, $^{\rm b)}$ broad and $^{\rm v)}$ very broad component (FWHM$>$8000 km s$^{-1}$) of the emission line.\\
$^{\rm m)}$  Marginal line detection ($2-3\sigma$),
$^{\rm \ast\ast)}$ Poor Gaussian fit ($\chi_{red} \sim 20$).\\
      \end{table*}
%

   \begin{table*}[t]
      \caption{FWHM of optical emission lines in [km s$^{-1}$], corrected for
      instrumental resolution.}
      \label{FWHM_opt}
      \begin{flushleft}
      \begin{tabular}{l@{\hspace*{3mm}}c@{\hspace*{3mm}}rrrrrrr}
      \hline\noalign{\smallskip}
name  & $z$ & [O II]~$\lambda$3727 & [Ne III]~$\lambda$3868 & H${\gamma}$~$\lambda$4340 & H${\beta}$~$\lambda$4861 & [O III]~$\lambda$4959 & [O III]~$\lambda$5007 & H${\alpha}$~$\lambda$6563 \\ \noalign{\smallskip}
(1) & (2) & (3)\hspace{0.6cm} & (4)\hspace{0.6cm} & (5)\hspace{0.4cm} & (6)\hspace{0.4cm} & (7)\hspace{0.7cm} & (8)\hspace{0.7cm} & (9)\hspace{0.4cm} \\ \noalign{\smallskip}
\hline\noalign{\smallskip}
18Z  &0.993& $<$420  & \point & \point & \point & \point & \point & \point \\
24Z  &0.480& 260$\pm$120  & \point & \point & 1440$\pm$60  & $<$320 & $<$410  & \point \\
33A  &0.974& 340$\pm$60  & \point & \point & \point & \point & \point & \point \\
34F  &0.267& 310$\pm$20  & \point & $<$470 & $<$370  & 120$\pm$50  & $<$420  & $<$350\\
34M  &0.267& $<$590  & \point & \point & $<$370  & $<$380  & $<$390  & $<$250  \\
\noalign{\smallskip}
70A  &1.008& 300$\pm$30  & \point & \point & \point & \point & \point & \point \\
104A &0.137& 370$\pm$140  & \point & \point & $<$360  & $<$510  & $<$410  & 190$\pm$10  \\
104C &0.134& $<$610  & \point & \point & $<$400  & $<$530  & $<$530  & 310$\pm$10  \\
128D &0.031& $<$750  & \point & $<$440  & $<$400  & $<$520  & $<$520  & $<$370  \\
426A &0.788& $<$520  & \point  & \point & \point & \point & \point & \point \\
\noalign{\smallskip}
513~(34O)  &0.761& $<670$  & 270$\pm$70 & \point & \point & \point & \point & \point \\
815D &0.694& 910$\pm$20  & \point & \point & \point & \point & \point & \point \\
815F &0.695& $<$540 & \point & \point & \point & \point & \point & \point \\
827B &0.249& \point & \point & \point & \point & \point & \point & 1170$\pm$10$^{*)}$  \\
870~(36F)  &0.807& 290$\pm$10  & 1440$\pm$60 & $<$320  & \point & \point & \point & \point \\
\noalign{\smallskip}
901A &0.205& 480$\pm$20  & 750$\pm$80  & \point & \point  & 600$\pm$50  & $<$600  & $<$480  \\
      \noalign{\smallskip}
      \hline
       \end{tabular}
      \end{flushleft}
$^{*)}$ A blend with emission lines [N II]~$\lambda\lambda 6548/6584$ cannot be excluded.

      \end{table*}
%
    
   \begin{table*}[t]
      \caption{Rest frame EW of optical emission lines in [\AA].}
      \label{EW_opt}
      \begin{flushleft}
      \begin{tabular}{l@{\hspace*{4mm}}c@{\hspace*{4mm}}rrrrrrr}
      \hline\noalign{\smallskip}
name &$z$ & [O II]~$\lambda$3727 & [Ne III]~$\lambda$3868 & H${\gamma}$~$\lambda$4340 & H${\beta}$~$\lambda$4861 & [O III]~$\lambda$4959 & [O III]~$\lambda$5007 & H${\alpha}$~$\lambda$6563 \\ \noalign{\smallskip}
(1) & (2) & (3)\hspace{0.6cm} & (4)\hspace{0.6cm} & (5)\hspace{0.4cm} & (6)\hspace{0.4cm} & (7)\hspace{0.7cm} & (8)\hspace{0.7cm} & (9)\hspace{0.4cm} \\ \noalign{\smallskip}
\hline\noalign{\smallskip}
18Z  &0.993& 3.7$\pm$0.9  & \point & \point & \point & \point & \point & \point \\
24Z  &0.480& 3.1$\pm$0.7  & \point & \point & 6.3$\pm$0.6  & 1.6$\pm$0.3  & 3.6$\pm$0.3  & \point \\
33A  &0.974& 7.2$\pm$0.9  & \point & \point & \point & \point & \point & \point \\
34F  &0.262& 57.6$\pm$2.4  & \point & 3.8$\pm$0.7 & 10.3$\pm$0.7  & 10.9$\pm$0.9  & 27.5$\pm$1.0  & 55.6$\pm$1.8  \\
34M  &0.262& 59.6$\pm$3.6  & \point & \point & 11.6$\pm$1.0  & 10.8$\pm$1.2  & 28.1$\pm$1.3  & 58.6$\pm$1.2  \\
\noalign{\smallskip}
70A  &1.008& 7.1$\pm$0.7  & \point & \point & \point & \point &\point & \point  \\
104A &0.137& 13.2$\pm$3.1  & \point & \point & 2.2$\pm$0.5  & 5.0$\pm$0.7  & 11.2$\pm$0.6  & 15.6$\pm$0.4  \\
104C &0.134  & 20.8$\pm$3.2  & \point & \point & 5.2$\pm$1.3  & 6.3$\pm$1.5  & 16.0$\pm$1.6  & 33.9$\pm$1.8  \\
128D &0.031& 43.3$\pm$2.8  & \point & 13.4$\pm$1.1  & 33.8$\pm$1.2  & 32.5$\pm$1.8  & 97.4$\pm$2.5  & 211.2$\pm$5.7  \\
426A &0.788& 7.7$\pm$1.3  & \point  & \point & \point & \point & \point & \point \\
\noalign{\smallskip}
513~(34O)  &0.761& 39.2$\pm$1.2  & 7.0$\pm$0.7 & \point & \point & \point & \point & \point \\
815D &0.694& 61.5$\pm$3.5  & \point & \point & \point & \point & \point & \point \\
815F &0.695& 45.1$\pm$2.1  & \point & \point & \point & \point & \point & \point \\
827B &0.249& \point & \point & \point & \point & \point & \point & 201.2$\pm$9.1$^{*)}$  \\
870~(36F)  &0.807& 15.9$\pm$0.4  & 4.3$\pm$0.4 & 2.3$\pm$0.2  & \point & \point & \point & \point \\
\noalign{\smallskip}
901A &0.205& 15.2$\pm$0.4  & 10.7$\pm$1.1  & \point & \point & 11.3$\pm$1.0  & 20.5$\pm$0.8  & 10.9$\pm$0.5  \\
      \noalign{\smallskip}
      \hline
       \end{tabular}
      \end{flushleft}
      \end{table*}
%


\begin{thebibliography}{}




  
  \bibitem[1976]{Av76} Avni Y., 1976, ApJ, 210, 642
  \bibitem[2000]{Aki00} Akiyama M., Ohta K., Yamada T., et al., 2000,
     ApJ 532, 700
  \bibitem[2001]{Ba01} Barger A.J., Cowie L.L., Mushotzky R.F., Richards E.A.,
     2001, AJ 121, 662
  \bibitem[1996]{Ber96} Bertin E., Arnouts S., 1996, A\&AS 117, 393
  \bibitem[1998]{Biz98} Bizenberger P., McCaughrean M. J., Birk C., et al.,
     1998, Omega Prime: the wide-field near-infrared camera for the 3.5m
     telescope of the Calar Alto Observatory. In A. M. Fowler, ed., Infrared
     astronomical instrumentation, SPIE vol. 3354, p. 825
  \bibitem[2000]{Bolz00} Bolzonella M., Miralles J., Pello R., 2000, A\&A (in press)
  \bibitem[1996]{Bow96} Bower R.G., Hasinger G., Castander F.J., et al.,
     1996, MNRAS 281, 59
  \bibitem[1995]{Boy95} Boyle B.J., McMahon R.G., Wilkes B.J.,
     Elvis M., 1995, MNRAS 272, 462
  \bibitem[1997]{Boy97} Boyle B.J., Wilkes B.J., Elvis M.,
     1997, MNRAS 285, 511
  \bibitem[2000]{Bra00} Brandt W.N., Hornschemeier A.E., Schneider D.P., et al., 2000, AJ 119, 2349
  \bibitem[1994]{Bro94} Brotherton M.S., Wills B.J., Steidel C.C., Sargent W.L.W.,1994, APJ 423, 131
  \bibitem[1983]{Bru83} Bruzual A.G., 1983, ApJ 273, 105
  \bibitem[1993]{Bru93} Bruzual A.G., Charlot S., 1993, ApJ 405, 538
  \bibitem[1999]{Cim99} Cimatti A., Daddi E., di Serego Alighieri S., et al., 
     1999, A\&A 352, L45
  \bibitem[1996]{Dav96} David F.R., Harnden Jr. F.R., Kearns K.E., et al., 
     1996, in: The ROSAT User Handbook, Briel et al., eds.
  \bibitem[1995]{Co95} Comastri A., Setti G., Zamorani G., Hasinger G.,
     1995, A\&A 296, 1
  \bibitem[1997]{deRui97} De Ruiter H.R., Zamorani G., Parma P., et al.,
     1997, A\&A 319, 7
  \bibitem[1996]{Geor96} Georgantopoulos I., Stewart G.C., Shanks T., et al.,
     1996, MNRAS 280, 276
  \bibitem[1962]{Giac62} Giacconi R., Gursky H., Paolini F.R., et al.,
     1962, Phys. Rev. Lett. 9, 439
  \bibitem[2000]{Gia00} Giacconi R., Rosati P., Tozzi P., et al., 2000,
     ApJ, in press (astro-ph/0007240)
   \bibitem[1999]{Gi99} Gilli R., Risaliti G., Salvati M., 1999,
      A\&A 347, 434
   \bibitem[2001]{Gi01} Gilli R., Salvati M., Hasinger G., 2001,
      A\&A 366, 407
  \bibitem[1996]{Gre96} Green P.J., 1996, ApJ 467, 61
  \bibitem[1987]{Gun87} Gunn J.E., Carr M.L., Danielson G.E., et al.,
     1987, Opt. Eng. 26, 779
  \bibitem[1998]{Has98} Hasinger G., Burg R., Giacconi R., et al.,
    1998, A\&A 329, 482 (Paper I)
  \bibitem[1999]{Has99b} Hasinger G., Giacconi R., Gunn J. E., et al.,
    1999b, A\&A 340, L27 (Paper IV)
  \bibitem[2001]{Has01} Hasinger G., Altieri B., Arnaud M., et al., 
    2001, A\&A 365, 45
  \bibitem[1986]{Hor86} Horne K., 1986, PASP 98, 609
  \bibitem[2000]{Horn00} Hornschemeier A.E., Brandt W.N., Garmire D.P.,
    et al., 2000, ApJ, in press (astro-ph/0004260)
  \bibitem[2001]{Ish01} Ishisaki Y., Ueda Y., Yamashita A., et al., 2001,
    submitted to PASJ
  \bibitem[1986]{Loc86} Lockman F.J., Jahoda K., McCammon D., 1986,
    ApJ 302,432
  \bibitem[1992]{LeB92} Le Borgne J.F., Pello R., Sanahuja B., 1992,
    A\&A Suppl. Ser. 95, 87
  \bibitem[2000a]{Leh00a} Lehmann I., Hasinger G., Schmidt M., et al.,
     2000a, A\&A 354, 34
  \bibitem[2000b]{Leh00b} Lehmann I., Hasinger G., Giacconi R., et al.,
     2000b, In the proceedings of the VLT Opening Symposium, held at Antofagasta
     (Chile), 1-4 March 1999, eds. J.Bergeron and A.Renzini, Springer-Verlag, p. 121
   \bibitem[1995]{Mad95}Madau, P., 1995, ApJ, 441, 18

   \bibitem[1994]{Ma94} 
      Matthews, K. \& Soifer, B.T. 
      1994, in {\em Experimental Astronomy}, 3, 77
  \bibitem[1998]{Mc98} McHardy I., Jones L.R., Merrifield M.R., et al.,
     1998, MNRAS 295, 641
  \bibitem[1996]{Mul96}Mulchaey J.S., Davis D.S., Mushotzky R.F., Burnstein D.,
     1996, ApJ 456, 80
  \bibitem[2000]{Mu00}Mushotzky R.F., Cowie L.L., Barger A.J., Arnaud K.A.,
     2000, Nat 404, 459
  \bibitem[2000]{Na00} Nakanishi K., Akiyama M., Otha K., Yamada T., 
     2000, ApJ 534, 587
  \bibitem[1983]{Oke83} Oke J.B., Gunn J.E., 1983, ApJ 266, 713
  \bibitem[1995]{Oke95} Oke J.B., Cohen J.G., Carr M., et al.,
     1995, PASP 107, 375
\bibitem[1998]{Pers98}
     Persson, S. E., Murphy, D. E., Krzeminski, W., Roth, M. \& Rieke, M. J. 
     1998, AJ, 116, 2745
  \bibitem[1994]{Pon94} Ponman T.J., Allan D.J., Jones L.R., et al., 1994,
     Nat 369, 462
 \bibitem[1992]{Pre92} Press H.W., Teukolski S.A., Vetterling W.T.,
    Flannery B.P., 1992b, Cambridge Press,
    Numerical Receipes in FORTRAN, Example Book, sec. edition
  \bibitem[1992]{Pfef82} Pfeffermann E., Briel U.G., 1982, IEEE Trans. Nucl.
     Sci., vol. 39, no. 4, p.976
  \bibitem[1997]{Puch97} Puchnarewicz E.M., Mason K.O., Carrera F.J.,
     et al., 1997, MNRAS 291, 177
  \bibitem[1999]{Ro99}Rosati P., Standford S.A., Eisenhardt P.R.,
     1999, AJ 118, 76
  \bibitem[1998]{Schm98} Schmidt M., Hasinger G., Gunn J.E., et al.,
     1998, A\&A 329, 495 (Paper II)
  \bibitem[1998]{Schn98} Schneider D.P., Schmidt M., Hasinger G., et al.,
     1998, AJ 115, 1230
  \bibitem[1991]{Stei91} Steidel C.C., Sargent W.L.W., 1991, ApJ 382, 433
  \bibitem[1994]{Stei94} Steidel C.C., Dickinson M., 1994, In Wide field
      spectroscopy and the distant universe, The 35th Herstmonceux Conference,
      held at Cambridge, United Kingdom, July 4-8, 1994, eds. S.J. Maddox and
      A. Aragon-Salamanca, World Scientific Publishing Co. Pte. Ltd., p. 349 
  \bibitem[1991]{Sto91} Stocke J.T., Morris S.L., Giola I.M., et al.,
     1991, ApJS 76, 813
  \bibitem[2001]{Tho01} Thompson D., et al., 2001, in prep.
  \bibitem[1987]{Vei87} Veilleux S., Osterbrock D.E., 1987, ApJS 63, 295
 \bibitem[1996]{Wil96} Wilson G., Kaiser N., Cole S., Frenk C., 1996,
     AAS 189, 8206
\bibitem[1999]{Zam99} Zamorani G., Mignoli M., Hasinger G., et al.,
   1999, A\&A 346, 731 (Paper V)
\end{thebibliography}
\end{document}